\documentclass[fleqn,usenatbib]{mnras}

\usepackage{newtxtext,newtxmath}
\usepackage{xcolor} 
\usepackage{placeins}

\usepackage[T1]{fontenc}
\usepackage{xcolor}

\DeclareRobustCommand{\VAN}[3]{#2}
\let\VANthebibliography\thebibliography
\def\thebibliography{\DeclareRobustCommand{\VAN}[3]{##3}\VANthebibliography}

\usepackage{graphicx}	% Including figure files
\usepackage{amsmath}	% Advanced maths commands
\usepackage{float}
\usepackage{subcaption}
\usepackage{siunitx}
\usepackage[normalem]{ulem}
\usepackage{mathtools}

\DeclareFontShape{OML}{cmm}{b}{it}{<->cmmi10}{}
\title[How massive neutrinos reshape the cosmic web]{How massive neutrinos reshape the cosmic web}

\author[Leonor N. L. Simões et al.]{Leonor N. L. Simões,$^{1}$\thanks{E-mail: leonor.simoes.20@ucl.ac.uk}
Krishna Naidoo,$^{2,1,3}$
Benjamin Joachimi,$^{1}$
Willem Elbers,$^{4}$
Carlos S. Frenk$^{4}$
\\
$^{1}$Department of Physics and Astronomy, University College London, Gower Street, London WC1E 6BT,  United Kingdom\\
$^{2}$Institute of Cosmology and Gravitation,University of Portsmouth, Burnaby Road, Portsmouth, PO1 3FX, United Kingdom\\
$^{3}$Max-Planck-Institut f\"{u}r Astronomie, K\"{o}nigstuhl 17, 69117 Heidelberg, Germany\\
$^{4}$Institute for Computational Cosmology, Department of Physics, Durham University, South Road, Durham, DH1 3LE, UK\\
}

\date{Accepted XXX. Received YYY; in original form ZZZ}

\pubyear{\the\year{}}

\begin{document}
\label{firstpage}
\pagerange{\pageref{firstpage}--\pageref{lastpage}}
    \maketitle

\begin{abstract}
We explore the effects of massive neutrinos on the cosmic web using the FLAMINGO simulations. We classify the cosmic web into voids, sheets, filaments, and clusters, and find that massive neutrinos affect the environment by decreasing the volume occupied by clusters and voids. We find that increasing the neutrino mass shifts the volume-weighted density distribution towards higher densities and leads to a more narrow density distribution, which we interpret as neutrinos delaying structure formation. We construct the minimum spanning tree (MST) graph from the subhaloes, adopting a number density chosen to match that expected for DESI-like observations. We show that most MST edges lie in filaments, approximately $70\%$ throughout different simulations, which we link to its sensitivity to neutrino mass. We also link the MST's edge length signal at different scales to different cosmic web environments, with clusters dominating the signal at small scales, voids at longer scales, and filaments at intermediate scales. The strong correlation between MST edges and cosmic web environments reinforces the MST’s potential to be used as a classifier for large-scale structure in galaxy surveys. We compare the effects of baryonic physics and massive neutrinos and find that each produces distinct signatures in MST edge lengths. This analysis is performed in 3D space, using the true positions of subhaloes and not accounting for redshift space distortions. Nevertheless, these results emphasise the MST's capability to go beyond two-point statistics, motivating future applications to real observational data.

\end{abstract}


%%%%%%%%%%%%%%%%%%%%%%%%%%%%%%%%%%%%%%%%%%%%%%%%%%

%%%%%%%%%%%%%%%%% BODY OF PAPER %%%%%%%%%%%%%%%%%%

\section{Introduction}
Over the next decade, observations of the large-scale structure of the Universe will advance our understanding of physics on the largest scales and subject the standard cosmological model to stringent tests. Current and future galaxy redshift surveys will determine the positions of hundreds of millions of galaxies, resulting in the deepest and largest map of the spatial distribution of matter in the Universe to date. These surveys include: the Dark Energy Spectroscopic Instrument (DESI, \citeauthor{2022DESI_overview} \citeyear{2022DESI_overview})\footnote{\href{https://www.desi.lbl.gov/}{https://www.desi.lbl.gov/}}, the European Space Agency's Euclid mission \citep{2025EuclidOverview}\footnote{\href{https://www.euclid-ec.org/}{https://www.euclid-ec.org/}}, and the Rubin Observatory Legacy Survey of Space and Time (LSST, \citeauthor{2019LSST_Overview} \citeyear{2019LSST_Overview})\footnote{\href{https://www.lsst.org/}{https://www.lsst.org/}}. Such surveys explore the Universe at later times, when the distribution of matter is highly non-Gaussian and ordered in a web-like  structure: the cosmic web \citep{bond1996filaments}.

Cosmologists have developed a series of methods for the purpose of cosmological parameter and model inference. Measurements of the Baryonic Accoustic Oscilations (BAO) from galaxy clustering \citep{alam2017clustering} and Lyman alpha absorption \citep{de2019baryon}, weak lensing \citep{abbott2018WL}, CMB anisotropies \citep{collaboration2020planck}, and standard candles, such as Ceipheid variables and type Ia supernovae \citep{riess2016candles}, are a few examples of these techniques. These probes have been widely used and studied, and thus have been consolidated as effective ways to establish and constrain the standard model of cosmology. 

Although these methods are reliable and trusted, most of them depend on measuring two-point statistics in both real (two-point correlation function, 2PCF) and Fourier space (the power spectrum). This is a very powerful tool to characterise the early universe (e.g. the cosmic microwave background), where almost all the information present in its highly Gaussian matter distribution is captured by the 2PCF. However, this technique fails to incorporate the highly non-Gaussian cosmic web information of the late universe. Recent studies have gone beyond two-point statistics and provided interesting constraints on cosmological parameters. These alternative statistical methods include N-point correlation functions \citep{slepian20173PFC, 2022NpointStats}, the bispectrum \citep{gil2016bspectrum}, wavelet scattering transforms \citep{2022Wavelet}, Minkowski functionals \citep{2025Minkowski}, density-split statistics \citep{2023densitysplit1, 2024densitysplit2}, artificial intelligence and machine learning \citep{Parroni2021}, among others.

The most attractive reason to explore methods that incorporate higher-order statistics is their potential to break existing parameter degeneracies, to provide tighter constraints and to test systematics. 
One fundamental parameter is the sum of the mass of the three neutrino species (which we will denote as $M_\nu$). The current lower bound for this mass comes from neutrino oscillation experiments: $M_\nu > 0.06 {\rm eV\, }$ \citep{Esteban24}, where we use natural units ($c=1$). The tightest upper bound, however, is given by cosmology \citep{DESI25, Camphuis25}, and is  $M_\nu < 0.05-0.07  {\rm eV\, }$. This result is achieved with the analysis of 2PCFs, in particular a combination of measurements of BAO with CMB data, and assuming a $\Lambda$CDM model. If the background cosmological model assumed is instead $w_0w_a$CDM, the upper constraint is more relaxed: $M_\nu < 0.163 {\rm eV\, }$. Moreover, the constraints from the free-streaming effect on the galaxy power spectrum are $M_\nu < 0.195 {\rm eV\, }$ \citep{Elbers_DESI_2025}. Although multiple analysis choices and data combinations yield different upper limits, large-scale structure observations are rapidly improving, pushing cosmology closer to a definitive measurement of the neutrino mass scale.

\cite{agarwal2011neutPP}, \cite{Daalen_2011}, \cite{Daalen_2015}, and \cite{schneider2019BarPP} showed that adding massive neutrinos to $\Lambda$CDM simulations and adding baryons to dark-matter-only simulations both show a similar suppression of the matter power spectrum on small scales ($k\approx10\text{Mpc}^{-1}h$). Moreover, \cite{Mummery_2017} showed that their impacts on several halo and clustering diagnostics are also qualitatively similar and can often be treated independently. In order to better constrain neutrino mass from observations, the effect of neutrinos should be distinguished from other effects like baryonic feedback \citep{2025FlamingoBaryonFeedback}, reiterating the need to go beyond the 2PCF and explore other techniques that are more sensitive to neutrino effects.

Studying the effect of massive neutrinos on cosmic web structures is also a key way to gain insights into their impact on structure formation. As shown in \cite{Bond80} and \cite{1998WeigheingNeutGalSur}, increasing the neutrino mass leads to a suppression of the growth of density perturbations, which makes the filamentary structure of the cosmic web more diffuse. This also results in shallower gravitational potential wells, and a redistribution of matter that increases the density in regions that would otherwise be underdense, such as voids. Since the cosmic web encodes rich information about both the growth of structure and the interplay between different matter components, understanding how neutrinos alter its morphology is crucial for constraining their properties and constraining cosmological models.

One promising technique to capture information from the cosmic web and estimate cosmological parameters in simulations and surveys is the minimum spanning tree (MST). In \cite{naidoo2020beyond}, the MST was shown to be a computationally inexpensive algorithm \citep{kruskal1956shortest} that can extract patterns from the cosmic web and go beyond two-point statistics. Additionally, \cite{naidoo21neutrino} showed that combining the MST with the traditional 2PCFs improved constraints on cosmological parameters by a factor of two and on neutrino mass by a factor of four. The origin of the MST's sensitivity to neutrino mass is unclear, but is believed to come from its ability to trace filaments, picking up on the effect of massive neutrinos washing out filamentary structures.

In this paper, we explore the relationship between the MST and the cosmic web to understand from where this sensitivity is coming. We use the FLAMINGO simulations \citep{FlamingoGeneral23, Kugel_2023}, a set of large hydrodynamical and dark-matter-only simulations with three neutrino masses: $0.06\,\mathrm{eV}$, $0.24\,\mathrm{eV}$, and $0.48\,\mathrm{eV}$. This makes them the most suitable choice to compare the effects of neutrinos with the effects of baryonic physics on the cosmic web environments and on the MST's statistics. We investigate whether the MST is able to break the degeneracy between the effects of neutrinos and baryonic physics seen in the power spectrum. Finally, we note that the power of the MST also lies in its ability to be used in galaxy surveys \citep{Naidoo24}, strengthening the idea that it is a promising tool for parameter inference in cosmology.

\section{Methods}
In this section, we describe the methods used to investigate the impact of massive neutrinos and baryonic physics on the cosmic web. We present  the simulations, the cosmic web classification approach, the MST and its statistics, and the error estimation procedure used in this paper.

\subsection{The FLAMINGO simulations}

The FLAMINGO (Full-hydro Large-scale structure simulations with All-sky Mapping for the Interpretation of Next Generation Observations) simulations are a large set of hydrodynamical simulations \citep{FlamingoGeneral23, Kugel_2023}. They are being used to study large-scale structure cosmology, galaxy formation and evolution, and neutrino cosmology (the focus of this paper).

In this paper, we use a subset of the full FLAMINGO suite, specifically focusing on simulations run in a box of length $L_{\mathrm{box}}=1\,\mathrm{Gpc}$, with $N_{\mathrm{c}}=1800^{3}$ cold dark matter particles and $N_{\nu}=1000^{3}$ neutrino particles. We study both the dark-matter-only plus neutrino (DMO $+\nu$) runs and hydrodynamical (HYDRO) runs simulated with $N_{b}=N_{c}$ baryon particles. The FLAMINGO simulations were computed with the hydrodynamical $N$-body simulation code \texttt{SWIFT} \citep{Schaller2024}, using the \texttt{SPHENIX} hydrodynamics scheme \citep{Borrow22} and the fast multipole and particle mesh methods for gravity calculations. The initial conditions were generated at $z=31$ with third-order Lagrangian perturbation theory using monofonIC \citep{Hahn20}, accounting for the separate transfer functions of cold dark matter and baryons \citep{Hahn21} and the presence of massive neutrinos \citep{Elbers22}. To ensure the simulations remain realistic for large-scale structure studies, the subgrid models for stellar and AGN feedback are calibrated to the observed low-redshift galaxy stellar mass function and cluster gas fractions \citep{FlamingoGeneral23}.
The simulations model baryonic physics using subgrid models for radiative cooling and heating \citep{Ploeckinger_2020}, star formation \citep{Schaye_2008}, stellar mass
loss \citep{Wiersma_2009, Schaye_2015}, supernova feedback energy \citep{Dalla_2008,Chaikin_2022,Chaikin_2023},
black hole seeding and growth, and thermal feedback from active galactic
nuclei \citep{Springel_2005,Booth_2009,Bahe_2022}. The subgrid parameters were calibrated using Gaussian process emulators \citep{Kugel_2023}.

We consider a subset of simulations assuming a Planck $\Lambda$CDM background cosmology \citep{collaboration2020planck} with different neutrino masses. Neutrinos were simulated using the $\delta f$ method \citep{Elbers2021} that provides extremely accurate predictions for neutrino physics even on very small scales. We use the DMO$+\nu$ and corresponding HYDRO simulations for three sets of cosmologies, where the neutrino mass is set to $0.06\,\mathrm{eV}$, $0.24\,\mathrm{eV}$, and $0.48\,\mathrm{eV}$. In Table \ref{tab:flamingo} we summarise the cosmological parameters of the specific FLAMINGO simulations used in this paper.

\begin{table*}
	\centering
	\caption{The cosmological parameters of the FLAMINGO simulations we use in this paper \citep{FlamingoGeneral23, Flamingo_neutrinos25}. For each simulation there is a gravity-only (DMO$+\nu$) run with combined cold dark matter (CDM) and baryon mass $m_{cb} = m_c + m_g$. The following columns are: the reduced Hubble constant ($h=H_0/100$, where $H_0$ is the present-day Hubble parameter in km/s/Mpc) in units of 100/s/Mpc; the present-day density of all matter ($\Omega_\text{m}$), cold-dark matter ($\Omega_\text{c}$), and baryons ($\Omega_\text{b}$); the sum of the mass of the neutrinos ($M_\nu$); the present root-mean-square matter fluctuation, averaged over a sphere of radius $8h^{-1}$Mpc ($\sigma_8$); the primordial normalisation of the power-spectrum ($A_s$); and the scalar spectral index ($n_s$)}
	\label{tab:flamingo}
	\begin{tabular}{lccccccccccc}
		\hline
        Simulation Cosmology & $m_\text{c}/\si{M_\odot}$ & $m_\mathrm{g}/\si{M_\odot}$ & $h$ & $\Omega_\text{m}$ & $\Omega_\text{c}$ & $\Omega_\text{b}$ & $M_\nu $ & $\sigma_8$ & $10^9 A_\text{s}$ & $n_\text{s}$ \\
		\hline
		{\footnotesize Planck}        & $5.72\times10^9$ & $1.07\times10^9$ & $0.673$ & $0.316$ & $0.265$ & $0.0494$ & $\SI{0.06}{\eV}$ & $0.812$ & $2.101$ & $0.966$  \\
		{\footnotesize Planck$\nu$0.24Fix} & $5.62\times10^9$ & $1.07\times10^9$ & $0.673$ & $0.316$ & $0.261$ & $0.0494$ & $\SI{0.24}{\eV}$ & $0.769$ & $2.101$ & $0.966$ \\
        {\footnotesize Planck$\nu$0.48Fix} & $5.52\times10^9$ & $1.07\times10^9$ & $0.673$ & $0.316$ & $0.256$ & $0.0494$ & $\SI{0.48}{\eV}$ & $0.709$ & $2.101$ & $0.966$ \\
    	\hline
	\end{tabular}
\end{table*}

The halo catalogue is computed using the Hierarchical Bound Tracing algorithm (HBT+, \citeauthor{HBT_halloes17} \citeyear{HBT_halloes17}, \citeauthor{Han_2018} \citeyear{Han_2018}, \citeauthor{Forouhar_2025} \citeyear{Forouhar_2025}), which exploits hierarchical structure formation to improve the identification of substructures compared to traditional halo finders. The catalogue is then processed by the Spherical Overdensity and Aperture Processor (SOAP,  \cite{McGibbon25}), which computes a large selection of (sub)halo properties in a range of apertures.

\subsection{Cosmic Web Classification with NEXUS+}

In the late-time universe, matter on large scales is arranged in a vast, web-like structure known as the cosmic web \citep{bond1996filaments}. The cosmic web is characterised by regions of high density called clusters, strands of matter with higher than average density called filaments, sheets of matter called walls, and underdense regions called voids. Cosmic web structures are often best described in the Zel'dovich approximation \citep{Zeldovich1970} as collapses along one axis forming a plane called a wall or a sheet, along two axes forming a line called a filament, and along all three axes forming a sphere called a node or cluster. Lastly, regions that exhibit no collapse along any dimension are defined as voids. This provides a precise definition for cosmic web structures, but in practice cosmic web structures are more amorphous and classifications tend to be highly dependent on the methodology being used (see \citealt{libeskind2018tracing} for a review of cosmic web classification schemes). This dependence is driven by the fact that cosmic web structures are defined by caustic structures, whose sharp boundaries are blurred out in realistic simulations or real data.

Motivated by cosmic structure formation from the Zel'dovich approximation, astronomers have often resorted to the Hessian matrix (i.e.~the three-dimensional second-order derivatives) of the density field \citep[T-web;][]{forero2009Tweb, Hahn_2007} or velocity field \citep[V-web;][]{Hoffman2012}. From the Hessian we can determine the number of collapsing dimensions (indicated with negative eigenvalues from the Hessian matrix) and thus determine the type of environment: node, filament, wall or void \citep{Doroshkevich_1970}. However, such a definition is dependent on the field being used, but in practice neither the continuous density nor the continuous velocity field is known precisely and instead has to be inferred and interpolated from discrete points (either particles from simulations or galaxies from real data).

This places a resolution dependence on the inferred cosmic web environments, which is typically resolved by first applying a Gaussian smoothing to the field prior to the computation of the Hessian matrix. This removes shot-noise from sampling the density or velocity field from discrete points and allows comparisons with different resolution simulations and/or data. However, the appropriate choice of the smoothing scale is not known a priori. Another choice is whether to smooth the density field or the logarithm of the density field. Since the density field is close to lognormal at late-times, logarithmic smoothing better preserves filamentary and wall-like structures but requires input density fields that do not contain empty cells.

To address the limitations of the T-web and V-web we use the NEXUS+ \citep{MMF2010, cautun2013nexus} cosmic web classification scheme. To allow logarithmic smoothing we either require the density field to be measured on a coarse grid via a particle mesh assignment scheme (such as cloud-in-cell) or to be obtained via the Delaunay tessellation field estimation (DTFE) technique \citep{Schaap2000}, ensuring non-zero density field estimations even in regions of low density. Grid-based particle mesh assignment schemes usually suffer strongly from shot noise effects in regions of low density, but this is typically only a limiting factor if the resolution of the simulation is relatively low in comparison to the resolution of the field. In this case, the DTFE method allows us to smoothly interpolate the density between particles.

In this paper, we use a cloud-in-cell particle mesh assignment scheme due to the resolution of the density field (900 voxels along each dimension) being relatively low compared to the particle resolution of the FLAMINGO simulations. For the hydrodynamical simulations, the density is computed by summing all particle types (i.e. dark matter, gas, stars and black holes). For the neutrino contribution to the density we measure the weighted sum of the particles before adding the neutrino average density following \citet{Elbers2021}.

To avoid an arbitrary choice of smoothing scale, in NEXUS+ we smooth the density field on five scales and measure the maximum response signature to each environment. The smoothing scales are $R_{n} = \big(\sqrt{2}\big)^{n}R_{0}$, with $R_{0}=1$ Mpc/h and measured across 5 smoothing scales with $n=0$ to $n=4$. For the cluster environments we simply smooth the density field, but for the filament and wall environments we use a logarithmic smoothing. The logarithmic smoothing is applied by transforming the field $f$ to
\begin{equation}
    g = \log_{10} f
\end{equation}
and smoothing with a Gaussian with scale $R_{n}$ to give us $g_{R_{n}}$ before transforming the field back to
\begin{equation}
    f_{R_{n}} = C_{R_{n}} 10^{g_{R_{n}}},
\end{equation}
where $C_{R_{n}}$ ensures the mean of $f$ is the same as $f_{R_{n}}$. The Hessian matrix of the density field at a given smoothing scale is then obtained from
\begin{equation}
    H_{ij,R_{n}}(\vec{x}) = R^{2}_{n}\frac{\partial^{2}f_{R_{n}}}{\partial x_{i}\partial x_{j}}(\vec{x}),
\end{equation}
where $i$ and $j$ are indices of the Hessian matrix, $R_{n}$ the Gaussian smoothing scale, $f_{R_{n}}$ the smoothed field and $\vec{x}$ the position on the field. The eigenvalues are sorted into ascending order: $\lambda_{1}\leq\lambda_{2}\leq\lambda_{3}$. We then obtain the shape strength
\begin{equation}
    \mathcal{I}_{R_{n}} = \begin{dcases}
        \,\,\bigg|\frac{\lambda_{3}}{\lambda_{1}}\bigg|, \quad &\mathrm{cluster},\\
        \,\,\bigg|\frac{\lambda_{2}}{\lambda_{1}}\bigg|\,\Theta\left(1-\bigg|\frac{\lambda_{3}}{\lambda_{1}}\bigg|\right), \quad &\mathrm{filament},\\
        \,\,\Theta\left(1-\bigg|\frac{\lambda_{2}}{\lambda_{1}}\bigg|\right)\,\Theta\left(1-\bigg|\frac{\lambda_{3}}{\lambda_{1}}\bigg|\right), \quad &\mathrm{wall},
    \end{dcases}
\end{equation}
and the cosmic web signature,
\begin{equation}
    \mathcal{S}_{R_{n}} = \mathcal{I}_{R_{n}}\times\begin{dcases}
        |\lambda_{3}|\,\theta(-\lambda_{1})\,\theta(-\lambda_{2})\,\theta(-\lambda_{3}), \quad &\mathrm{cluster},\\
        |\lambda_{2}|\,\theta(-\lambda_{1})\,\theta(-\lambda_{2}), \quad &\mathrm{filament},\\
        |\lambda_{1}|\,\theta(-\lambda_{1}), \quad &\mathrm{wall}.
    \end{dcases}
\end{equation}
where $\theta(x)$ is the Heaviside function, defined as $1$ if $x\geq0$ and zero otherwise, and $\Theta(x) = x\,\theta(x)$. The final signature is taken to be the maximum signature across all smoothing scales,
\begin{equation}
    \mathcal{S} = \max_{\mathrm{levels}\,\,n} \mathcal{S}_{R_{n}}.
\end{equation}

We follow \citet{cautun2013nexus} in obtaining cluster, filament, and wall signature thresholds. For clusters, the signature threshold is taken to be the value at which half the cluster environments with a mass above $5\times10^{13}M_{\odot}\,h^{-1}$ have densities above $300$ times the normalised density. For filaments and walls, the signature threshold is taken to be the peak of the derivative of the mass change as a function of the signature threshold. To remove spurious environment signals we also filter out environments that have a volume less than $0.5\,h^{-3}\mathrm{Mpc}^{3}$ for clusters and $10\,h^{-3}\mathrm{Mpc}^{3}$ for filaments and walls. The NEXUS+ cosmic web classifications were computed with the Cosmic-web Analysis and Classification Toolkit for Unveiling Structure (\texttt{CACTUS}; Naidoo et al.~in prep.).

\begin{figure*}
     \centering
    \includegraphics[width=\textwidth]{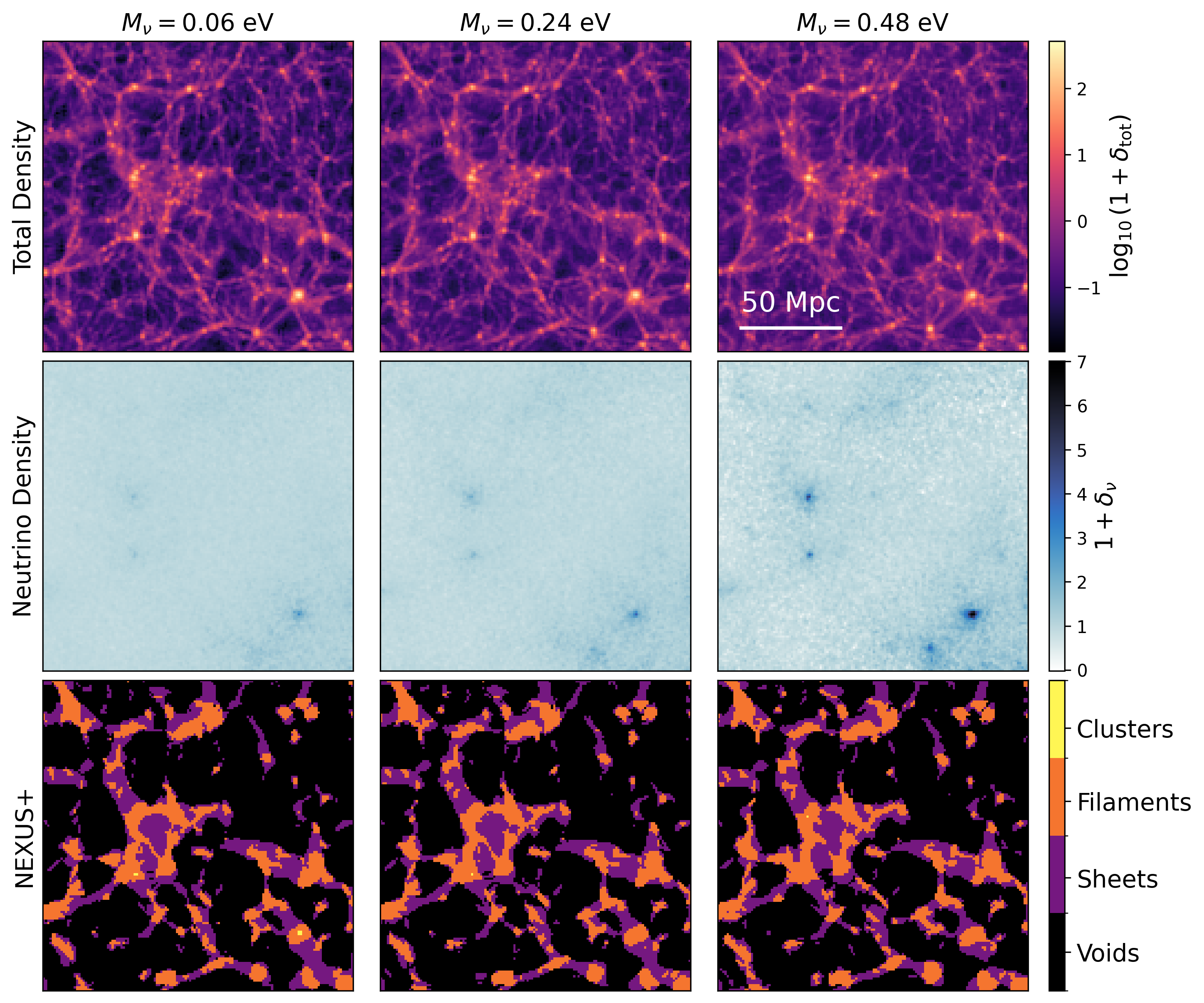}
        \caption{The same slice (projection of $155\times 155\times 5 \text{Mpc}^{3}$ subvolume) of the total density contrast (top row, where $1+\delta_{total} = \frac{\rho_{total}}{\bar{\rho_{total}}}$), the neutrino density contrast (centre row, where $1+\delta_\nu = \frac{\rho_\nu}{\bar{\rho_\nu}}$), and the NEXUS+ classification (bottom row) for each simulation with different neutrino masses (each column is one mass). This highlights the subtlety of neutrino effects: although in the centre row we can see the neutrino density increasing, the differences throughout the top and bottom rows are minimal.}
        \label{fig:contrast}
\end{figure*}

In Fig. \ref{fig:contrast} we show the same two-dimensional slice over the three DMO$+\nu$ simulations with the different neutrino masses. We plot the density field, the neutrino density, and the cosmic web classification from NEXUS+. It is clear that the effects of neutrinos on the cosmic web are subtle: as we increase neutrino mass we see a very slight increase of filaments and sheets taking over the space of voids. We see this in the structures on the bottom right side of the NEXUS+ panel.

We note that, while NEXUS+ is a physically motivated and consistently applied scheme, the classification of voxels into cosmic web environments inevitably involves choices, such as the smoothing scales and signature thresholds, that are not uniquely defined. However, varying these choices within reasonable bounds leaves our main results unchanged, indicating that our conclusions are not driven by a specific NEXUS+ implementation.

\subsection{The Minimum Spanning Tree (MST)}

The MST is a highly optimised graph that reduces a fully connected graph to a tree (a graph with no loops) that is spanning (connecting all nodes in a single structure) and has the minimal total length. The construction of this graph has been shown to be very good at picking out topological features, and in particular filaments \citep{Bhavsar1988, bhavsar1988filamentsreal, colberg2007quantifying, Alpaslan2014, libeskind2018tracing}, from a point data set.

In fact, the MST has been used in the field of astronomy for cosmic web filament finding in the galaxy  distribution since the work of  \citet{barrow1985minimal}.
In \citet{naidoo2020beyond} it was shown how to extract statistics from the MST for cosmology and in \citet{naidoo21neutrino} the MST was shown to greatly improve the constraints on cosmological parameters and neutrino mass relative to 2PCFs, a factor of 2 improvement in all parameters and a factor of 4 improvement in neutrino mass constraints.

Each MST will have a total of $n-1$ edges, with $n$ being the number of nodes \citep{kruskal1956shortest}. Following \citet{naidoo2020beyond}, we measure the distribution of the following MST statistics:

\begin{itemize}
    \item Degree ($d$) - the number of edges connected to each point.
    \item Edge length ($l$) - the length of an edge; which is taken to be the Euclidean distance between the end points.
    \item Branch length ($b$) - the sum of the length of branch member edges, where branches are edges connected along chains (i.e. points with degree $d=2$; \citealt{rainbolt2017use}).
    \item Branch shape ($s$) - a measure of the straightness of a branch, taken to be the straight line distance between the end points of a branch divided by the branch length $b$.
\end{itemize}
The MST's unique ability to extract patterns from points has been utilised for a wide and diverse set of scientific research and scientific disciplines. This includes applications in social sciences \citep{chang2011classification}, network topology \citep{macdonald2005minimum}, computer science \citep{chin1978CS_MST}, and astronomy \citep{graham1995quasarMST}. 
In this paper, the MST was constructed and analysed using the public python package \texttt{MiSTree} \citep{Naidoo2019Mistree}, using subhaloes as nodes.

\subsection{Error estimation: Jackknife resampling}
All errors in this paper were estimated using a cross-validation 
jackknife resampling technique.
Given a sample of size $n$, jackknife estimates of the mean and variance can be obtained by omitting a single observation and recomputing the statistic of interest $n$ times \citep{miller1974jackknife}. In practice, one does not necessarily need to divide by the sample size, and instead the overall data set can be partitioned into $n$ sets, where each jackknife sample is computed with one of these partitions omitted.

For the statistics measured in this paper, we partition the single simulation box into $n$ subboxes of equal size (where $n=5^3$: a cube number for $5$ equal divisions along each axis). We then calculate the jackknife mean:
\begin{equation}
    {\displaystyle {\bar {x}}_{\mathrm{JK} }={\frac {1}{n}}\sum _{i=1}^{n}{\bar {x}}_{(i)}.}
\end{equation}
where $\bar {x}_{(i)}$ is the summary statistic for the jackknife sample (i). 
We can then obtain the standard deviation using the expression:
\begin{equation}
    {\displaystyle {{\operatorname \sigma }}({\bar {x}})_{\mathrm {JK} }=\sqrt{{\frac {n-1}{n}}\sum _{i=1}^{n}({\bar {x}}_{(i)}-{\bar {x}}_{\mathrm {JK} })^{2}}}
\end{equation}

\section{Results}
\subsection{How do neutrinos affect the cosmic web structures?}\label{sec:model_affects}

In this section, we examine the effects of neutrino mass on the cosmic web environments by looking at how it affects the volume and mass fractions as well as the overall density distributions. The aim is to investigate what happens directly on the matter fields when we add heavier neutrinos, and then link this to what we measure indirectly in Section \ref{sec:MST analysis}.

\subsubsection{Global Statistics}
\begin{figure}
    \centering
    \includegraphics[width=\columnwidth]{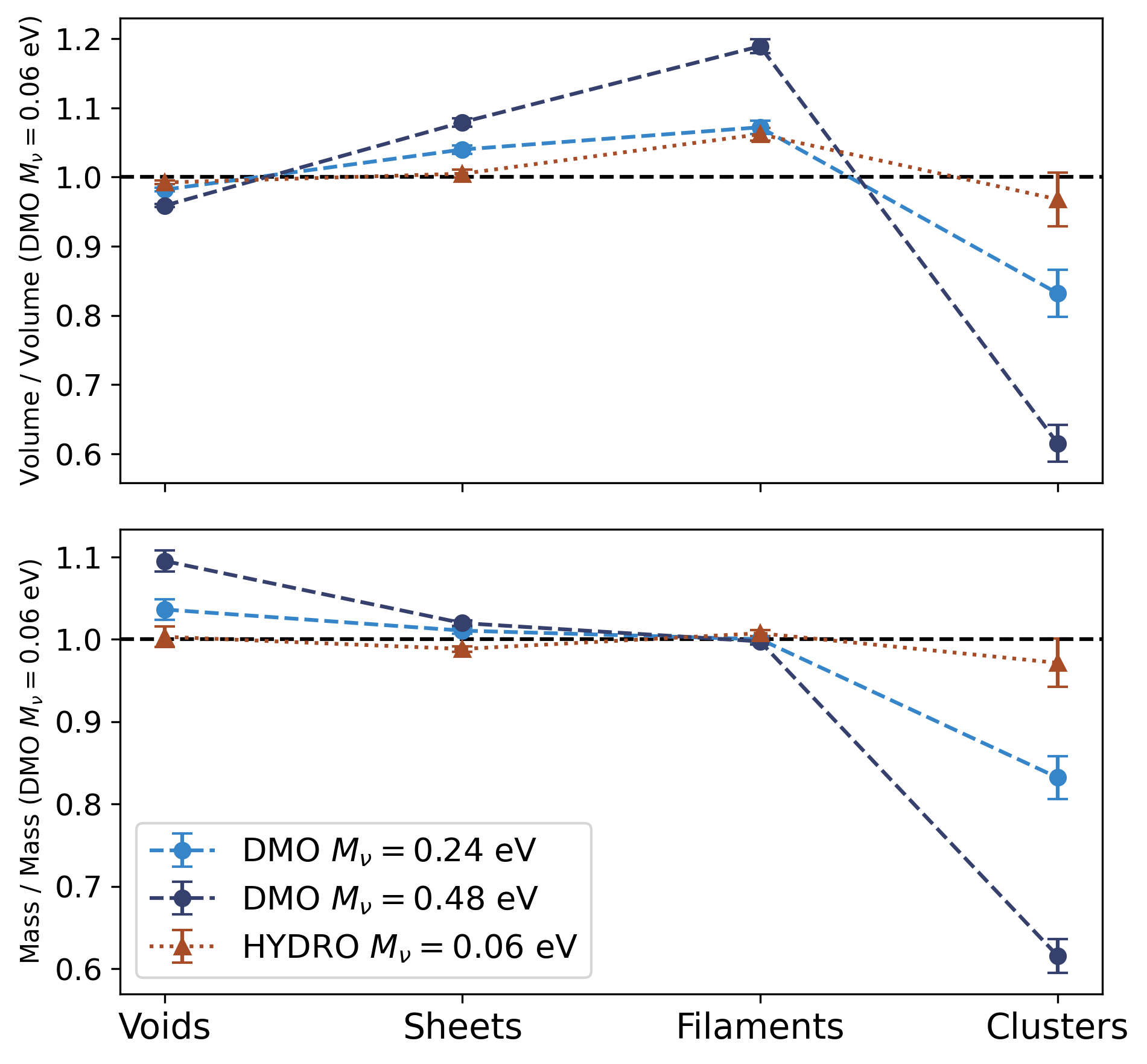}
    \caption{Volume (top) and mass (bottom) fractions, relative to the values for the DMO$+\nu$ simulation with $M_\nu = 0.06\,\mathrm{eV}$.}
    \label{fig:fraction}
\end{figure}

In Fig. \ref{fig:fraction} we plot the volume and mass fraction of each cosmic web component relative to the values for the DMO$+\nu$ simulation with $M_\nu = 0.06 \text{ eV}$. This allows us to isolate the effects of massive neutrinos and baryons on the global statistics of the cosmic web.

First, we notice that adding baryons to the simulation has a small effect on the environments. We distinguish an increase in filaments by volume, a small decrease in sheets by mass, and a decrease in clusters by both volume and mass (although this cluster effect is within the error bars it is still a feature we pick up). Baryonic effects are known to lower the mass of clusters by processes such as AGN feedback \citep{Mummery_2017, Semboloni_2011, Semboloni_2013}, which could explain the decrease in clusters. The subsequent increase in filaments by volume is  a consequence of AGN feedback ejecting gas at high redshifts ($z \approx 2-4$), during the epoch of rapid black hole growth \citep{McCarthy_2011}, which slows the subsequent growth of structure in a manner qualitatively similar to the effect of massive neutrinos \citep{Mummery_2017}. We note that recent kSZ measurements suggest the fiducial FLAMINGO feedback model may underestimate the true degree of gas ejection \citep{McCarthy_2025}, which could further amplify this effect.
%The subsequent increase in filaments by volume could be due to a redistribution of matter from the cluster feedback, causing the filaments to expand.

Let us focus now on the effects of massive neutrinos. We see a decrease in clusters by both mass and volume, which we can link to neutrinos suppressing the collapse of structure and washing out power on small scales, which leads to an exponential suppression at the heavy end of the halo mass functions \citep{Brandbyge10}. This suppression also leads to an increase in filaments and sheets by volume, as the matter that would have collapsed into clusters instead lingers in less dense, more extended structures. As filaments and sheets expand to fill more space, the volume occupied by voids decreases, but the matter is washed out into such voids making them more massive in general: in other words, voids become shallower. All this shows how increasing neutrino mass alters the cosmic web by suppressing high-density structures and enhancing the volume of intermediate and diffuse environments, leading to a smoother matter distribution.

\subsubsection{Density statistics}
We examine the effect of neutrino mass on the overall matter distribution as well as split into  different NEXUS+ cosmic web environments.

\begin{figure}
        \includegraphics[width=\columnwidth]{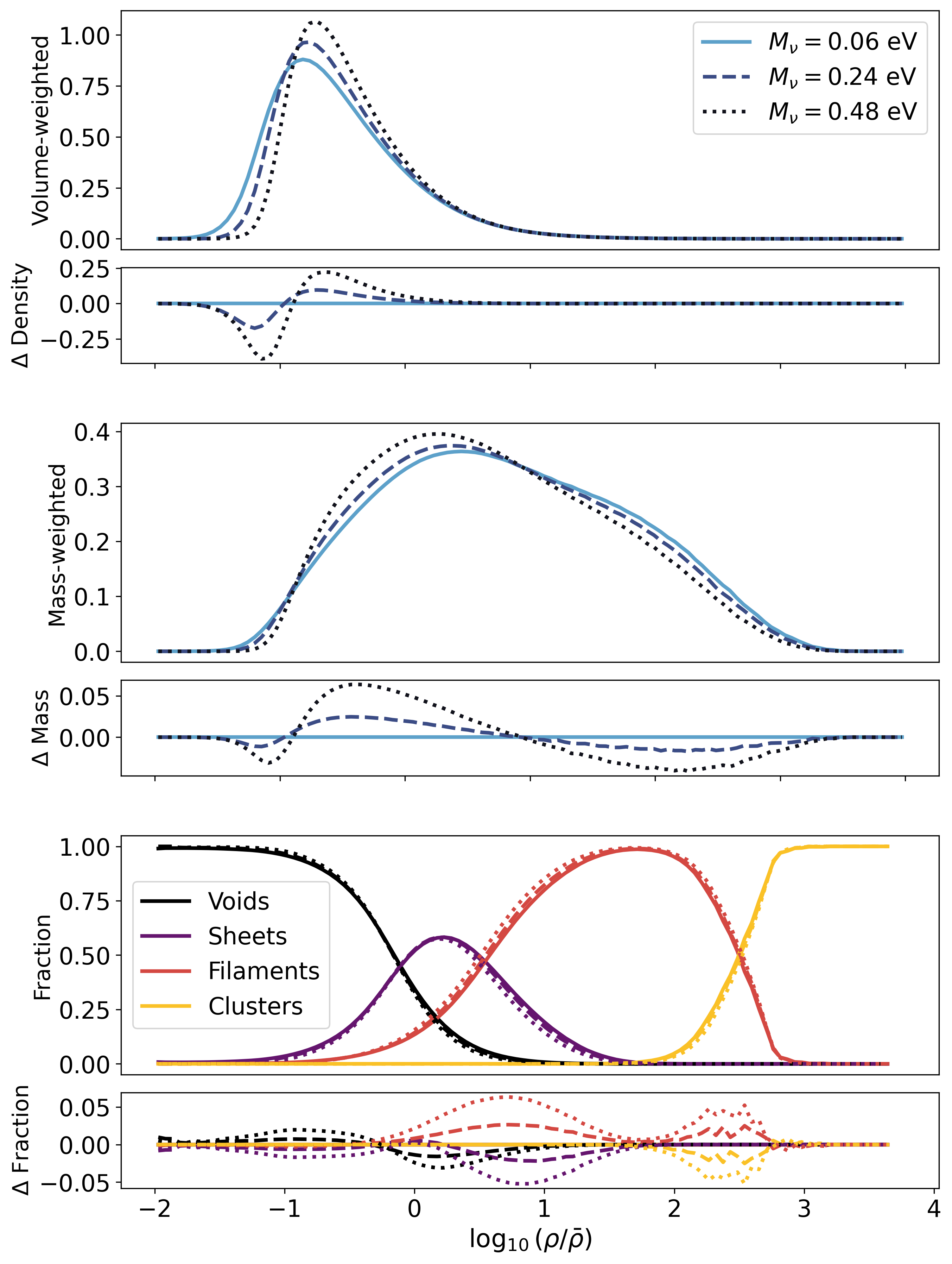}
        \caption{The normalised density (top) and mass-weighted density (middle) distributions for each neutrino mass. The top panels show the total density whereas in the bottom panels we subtract the $M_{\nu}=0.06\:{\rm eV\, }$ distribution from the other two. In the bottom panel we split by cosmic web environment to show which components dominate at different densities.}
        \label{fig:density_models}
\end{figure}

In Fig. \ref{fig:density_models} we plot the volume and mass-weighted density distributions in each DMO$+\nu$ simulation, comparing between the three neutrino masses and split by the cosmic web environment. The volume-weighted density distribution quantifies the fraction of volume at each given density biy. In contrast, the mass-weighted distribution measures the fraction of total mass contained in regions of a given density, thereby up-weighting high-density environments.The top panels of this figure highlight that increasing the neutrino mass shifts the peak of the volume-weighted density distribution towards higher densities and decreases the width of the overall distribution. These effects can be interpreted as neutrinos delaying structure formation: an increasing average density suggests there are less deep voids; while a narrower density distribution implies an environment with less dense clusters and less empty voids. This suggests that voids are also sensitive to neutrino effects and can be a powerful tool to constrain their mass \citep{PowerofCosmicWeb25Sunseri}. In the mass-weighted density distribution we see a similar effect plus an additional density loss at large densities when we increase neutrino mass, which we link again to the neutrinos washing out power on small scales.

In the bottom panel of Fig. \ref{fig:density_models} we see a hierarchy of classification with respect to density: voids dominate at lower densities, then come sheets and filaments, and finally clusters at higher densities. This is by construction. Notice how increasing neutrino mass decreases the width of the distribution of sheets and increases the width of the filament distribution. 

To see this interchange more clearly we focus on the smaller panels in Fig. \ref{fig:density_models} that show the difference with respect to $M_\nu =0.06\,{\rm eV\, }$. These plots highlight how increasing the neutrino mass affects the NEXUS+ classification at different densities. At low densities, we see an increase in voids at the cost of a decrease in sheets, reinforcing the voids' sensitivity to massive neutrino effects. Another interesting result is that we see a clear increase in filaments at medium to high densities, at the cost of a combination of voids and sheets at medium densities and clusters at higher densities. This highlights the filament's sensitivity to neutrino mass. Furthermore, filament environments dominate quite a large range in density. This demonstrates how, although much of the volume is dominated by the void regions, the mass is dominated by filaments, which is why filaments are more sensitive to the cosmic web and neutrino signals in general.

We note that we do not plot the effects of baryons because when we tested the effects of baryonic physics on density distributions, we found it to be negligible.

\subsection{Minimum Spanning Tree analysis} \label{sec:MST analysis}

To construct the MST, we choose from the more massive dark matter subhaloes, imposing a lower mass cut of $ 10^{12} M_\odot$ . We do this to remove the MST's sensitivity to the density tracer. This removes the sensitivity to particle mass resolution and enables comparisons between simulations with different particle mass resolutions (for example comparison between DMO$+\nu$ and HYDRO simulations). Additionally, the more massive subhaloes correspond to regions that are more likely to host the formation of galaxies, which are observable. 

We then pick 500 000 subhaloes at random to act as nodes from which we construct the MST. This corresponds to a density of approximately 0.0016 galaxies per $\text{Mpc}^{-3} \text{h}^{3}$, which is consistent with current and upcoming galaxy surveys \citep{galaxy_density23} such as DESI. We are selecting the same matched subhaloes for DMO$+\nu$ and Hydro simulations with the same neutrino mass and the mass cut is performed on the DMO$+\nu$ simulation.

\subsubsection{How does the MST trace the cosmic web?} \label{sec:edgeslie}
To better understand how the MST traces large-scale structure, we investigate where its edges lie within the cosmic web. By identifying the cosmic web environment of each edge, we can assess how the MST relates to the underlying distribution of voids, sheets, filaments, and clusters. This analysis provides insight into the sensitivity of the MST to the different cosmic web structures by determining in which environments its edges are embedded.

\begin{figure*}
     \centering
     \begin{subfigure}[b]{0.49\textwidth}
         \centering
         \includegraphics[width=\textwidth]{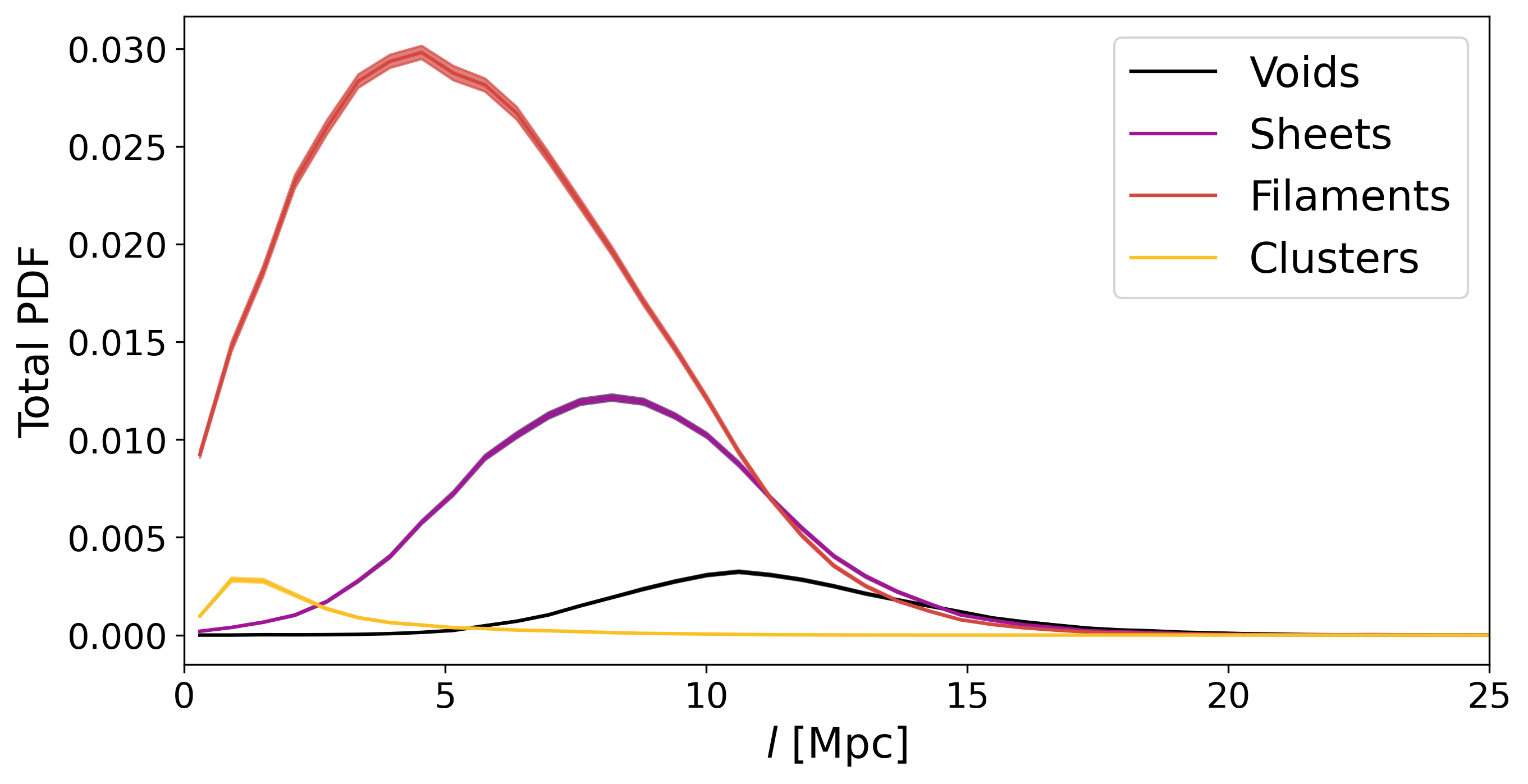}
         \caption{}
         \label{fig:neut_edges}
     \end{subfigure}
     \hfill
     \begin{subfigure}[b]{0.49\textwidth}
         \centering
         \includegraphics[width=\textwidth]{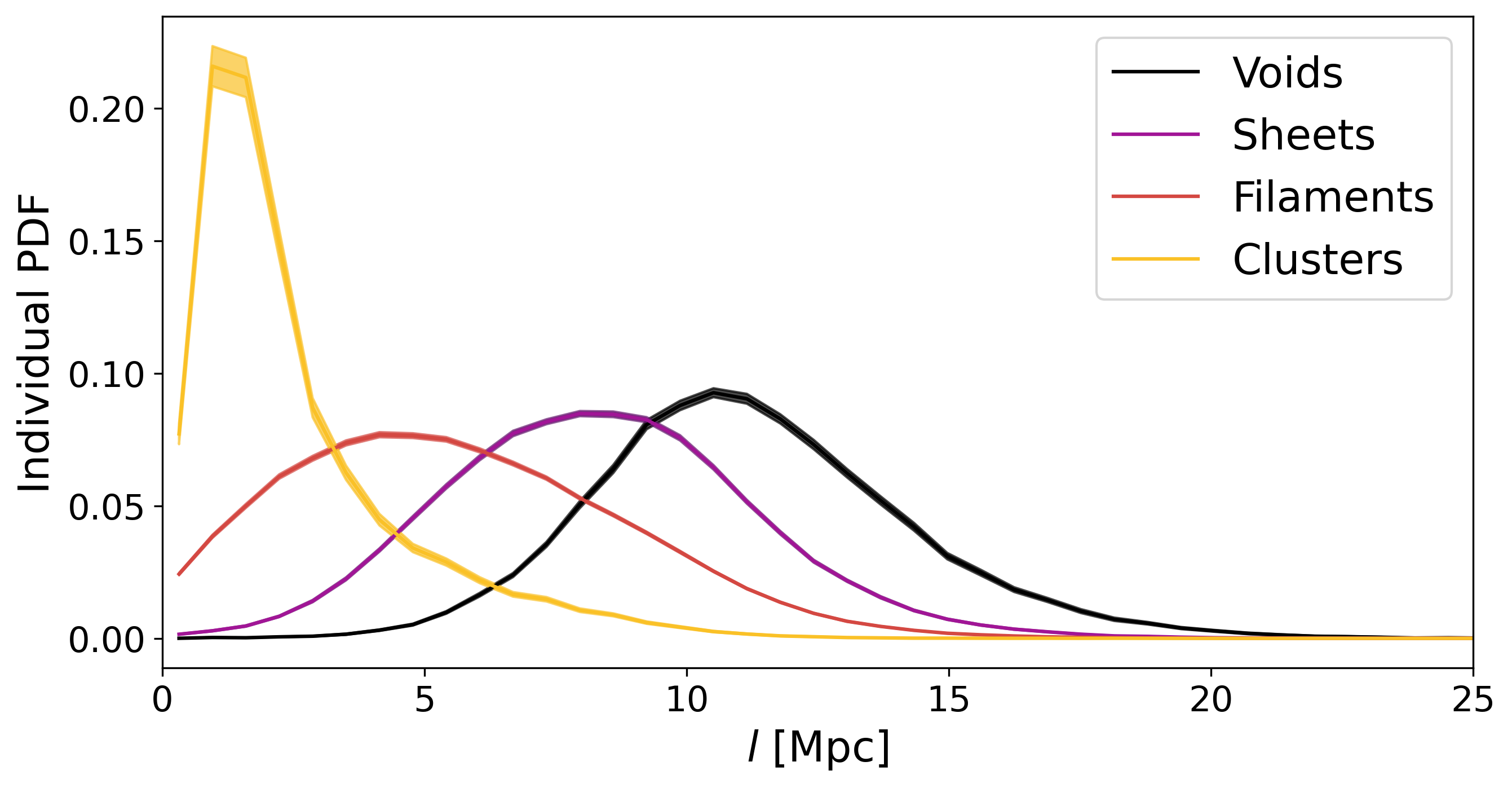}
         \caption{}
         \label{fig:neut_PDF_edges}
     \end{subfigure}
        \caption{Histograms of the distribution of the edge lengths ($l$) in the MST separated into each classification of the cosmic web that they are tracing. In Fig. \ref{fig:neut_edges} we plot the total probability distribution function (PDF), normalised to total counts. In Fig. \ref{fig:neut_PDF_edges} we plot the individual PDFs, normalised for each classification. These plots come from the analysis of the DMO$+\nu$ simulation with $M_\nu = 0.06 \:{\rm eV\,}$, but the relations are similar for all simulations.}
        \label{fig:edges_neutrinos}
\end{figure*}

Finding where a single point is in the cosmic web environment is straightforward since we just find what environment is associated to its coordinate. However, edges are lines that connect two nodes, and so to find out where the MST edges are in the NEXUS+ classification, we need to account for the possibility that the same edge can be going through different environments. To do this for all edges, we find the longest edge in each MST ($l_{\max}$) and divide it by the minimum resolution of the NEXUS+ classification in units of Mpc. We then round this value up (to make sure we oversample) to the nearest higher integer to get the number of equal parts we want to split each edge into. We split every edge into this number of parts. This ensures that we account not only for the fact that edges might probe multiple environments but also for the statistical weight associated with each environment. For example, if we split one edge into $5$ equal parts, we might find $2/5$ of them are in a filament and $3/5$ in a sheet. We can then associate each part of the edges with a given classification (void, sheet, filament, or cluster). In Fig. \ref{fig:edges_neutrinos} we show the relationship between the distribution of original edge length ($l$) and the cosmic web environment.

It is clear from Fig. \ref{fig:edges_neutrinos} that the MST is very good at picking out filamentary structures from the NEXUS+ classification (Fig. \ref{fig:neut_edges}). We can link this to how the MST is constructed: it finds an optimal route for connecting points in space, so naturally filaments emerge, since they represent connections between the different structures of the cosmic web.

In Table \ref{tab:edgeslie} we show where the MST edges lie for the different simulations. This table shows that both increasing neutrino mass and adding baryonic physics have a similar effect: decreasing the MST edges in voids and sheets and increasing the ones in filaments. The majority of the edges ($\approx 70 \%$ over all simulations) lie in filaments, which suggests that the MST is sensitive to the properties and changes of filamentary structure.

\begin{table}
    \centering
    \caption{Percentages of where the MST edges lie in the NEXUS+ classification.}
    \label{tab:edgeslie}
    \begin{tabular}{lcccc}
        \hline
        Simulation\_$M_{\nu}$ & Voids & Sheets & Filaments & Clusters \\
        \hline
        DMO\_0.06 & 6.01 & 24.45 & 67.27 & 2.28 \\
        DMO\_0.24 & 5.01 & 22.21 & 70.60 & 2.18 \\
        DMO\_0.48 & 4.78 & 21.64 & 72.04 & 1.54 \\
        HYDRO\_0.06 & 5.12 & 22.03 & 70.37 & 2.48 \\
        HYDRO\_0.24 & 5.22 & 22.38 & 70.47 & 1.92 \\
        HYDRO\_0.48 & 4.26 & 19.88 & 74.23 & 1.63 \\
        \hline
    \end{tabular}
\end{table}
 
Another interesting result is that there is a hierarchy of classifications, shown in Fig. \ref{fig:neut_PDF_edges}. Shorter edge lengths pick out clusters and filaments, then at longer lengths we see a peak in sheets, followed by voids at even longer lengths. This makes sense since we can think of the length of the MST edges as a measure of the distance between adjacent dark matter haloes, and we expect the filaments and clusters to be more dense regions (hence haloes are closer together) than sheets and voids. As we saw before in Section \ref{sec:model_affects}, this hierarchy is also present when we plot the density distribution in each cosmic web classification. The fact that a similar sensitivity appears in other graph-based constructions, such as Delaunay triangulation \citep{Dakshesh_25}, suggests that these methods consistently capture physically meaningful structures and can be powerful tools for extracting this information from real data.

It is worth noting that we do not see this sort of behaviour by choosing pairs of haloes at random to create edges, as opposed to the ones picked by the MST. In the random case we see the distributions follow a volume effect: the edge fraction grows with edge length for all environments since the probability of choosing two haloes close together is smaller than two further apart.

\subsubsection{MST statistics}

\begin{figure*}
     \centering
     \begin{subfigure}[b]{0.95\textwidth}
         \centering
         \includegraphics[width=\textwidth]{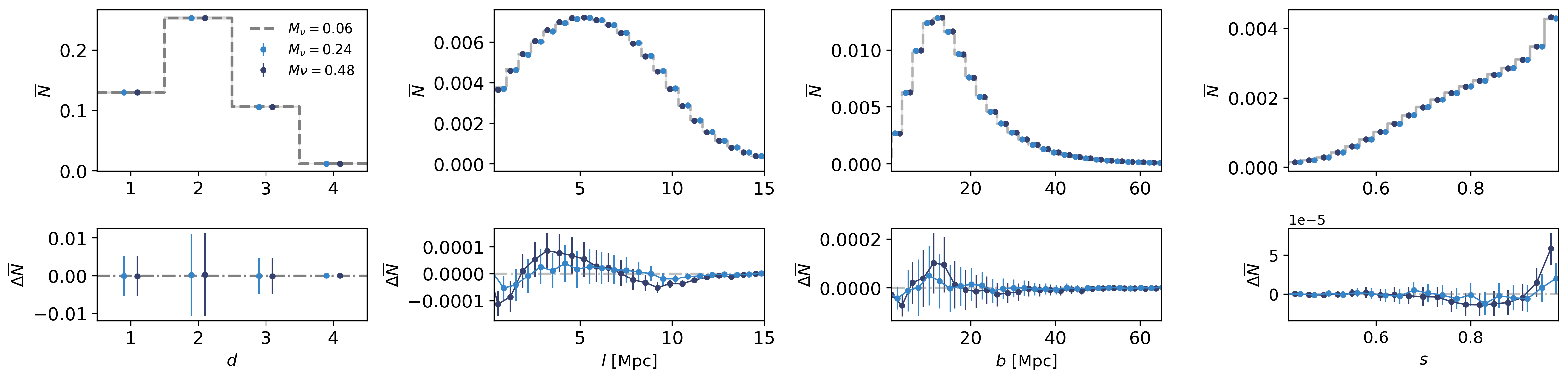}
         \caption{}
         \label{fig:neut_MST}
     \end{subfigure}
     \hfill
     \begin{subfigure}[b]{0.95\textwidth}
         \centering
         \includegraphics[width=\textwidth]{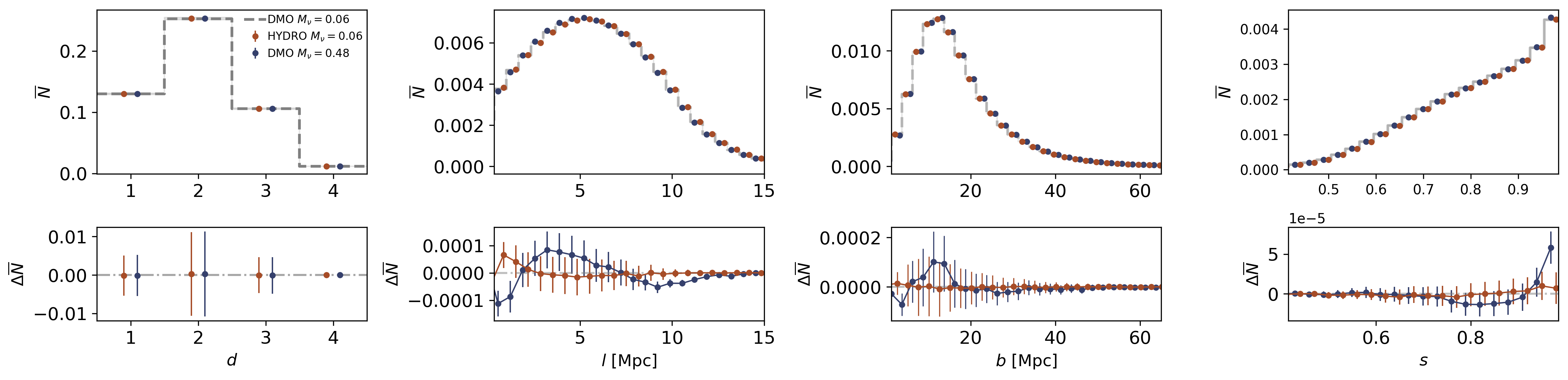}
         \caption{}
         \label{fig:Baryon_MST}
     \end{subfigure}
        \caption{A comparison between the histogram distributions of the MST statistics in different models, using 500 000  haloes for the MST. In Fig. \ref{fig:neut_MST} we compare different neutrinos masses, and in Fig. \ref{fig:Baryon_MST} we compare the effect of baryons and the effect of massive neutrinos. In the top panels we show the distributions of the degrees (\textit{d}), edge lengths (\textit{l}), branch lengths (\textit{b}), and branch shapes (\textit{s}). The bottom panels show the differences between the PDFs, subtracting the initial $M_\nu = 0.06\:{\rm eV\, }$ distribution (the dashed black line shows zero) from each model: $\Delta\bar{N}=\bar{N}_{model} -\bar{N}_{DMO,M_{0.06}}$.}
        \label{fig:mst_comp}
\end{figure*}

We plot and compare the MST statistics for each simulation in Fig. \ref{fig:mst_comp}. In Fig. \ref{fig:neut_MST} we compare the effect of neutrino mass on the MST statistics. In this figure we concentrate on the DMO$+\nu$ simulations, but similar results were found for the HYDRO simulations.

Adding massive neutrinos decreases the number of edge lengths at the smallest scales of $l \approx 1 \,\mathrm{Mpc}$ , increases the small/intermediate scales of $l \approx 3\,\mathrm{Mpc}$, and decreases the number of edge lengths at larger scales of $l \approx 9\,\mathrm{Mpc}$. It is also clear that increasing the mass of the neutrinos increases the magnitude of this effect.

In Fig. \ref{fig:Baryon_MST}, we distinguish the effects of massive neutrinos from the effects of baryons. We use a DMO$+\nu$ simulation with the smallest neutrino mass as the baseline, which we compare to a HYDRO simulation with the same neutrino mass and a DMO$+\nu$ simulation with the neutrino mass set to $M_\nu = 0.48\:{\rm eV\,}$. The two effects are distinguishable in the edge length panel, with the rest of the statistics proving to be more similar and noisy. Baryons increase the number of edges at the smallest scales of $l\approx 1\,\mathrm{Mpc}$ and show a similar edge length distribution to DMO$+\nu$ in the following lengths. This is the opposite of the neutrino effect at small scales, and both effects have a unique signature. 

The MST's capability to distinguish the effects of baryons and neutrinos means it could be a useful statistic for breaking degeneracies in traditionally used two-point statistics. Furthermore, this unique signal in edge length is most likely the reason why the MST is so sensitive to neutrino mass, as seen in \citet{naidoo21neutrino}.

\subsubsection{MST statistics in cosmic web environments}

\begin{figure}
         \centering
         \includegraphics[width=\columnwidth]{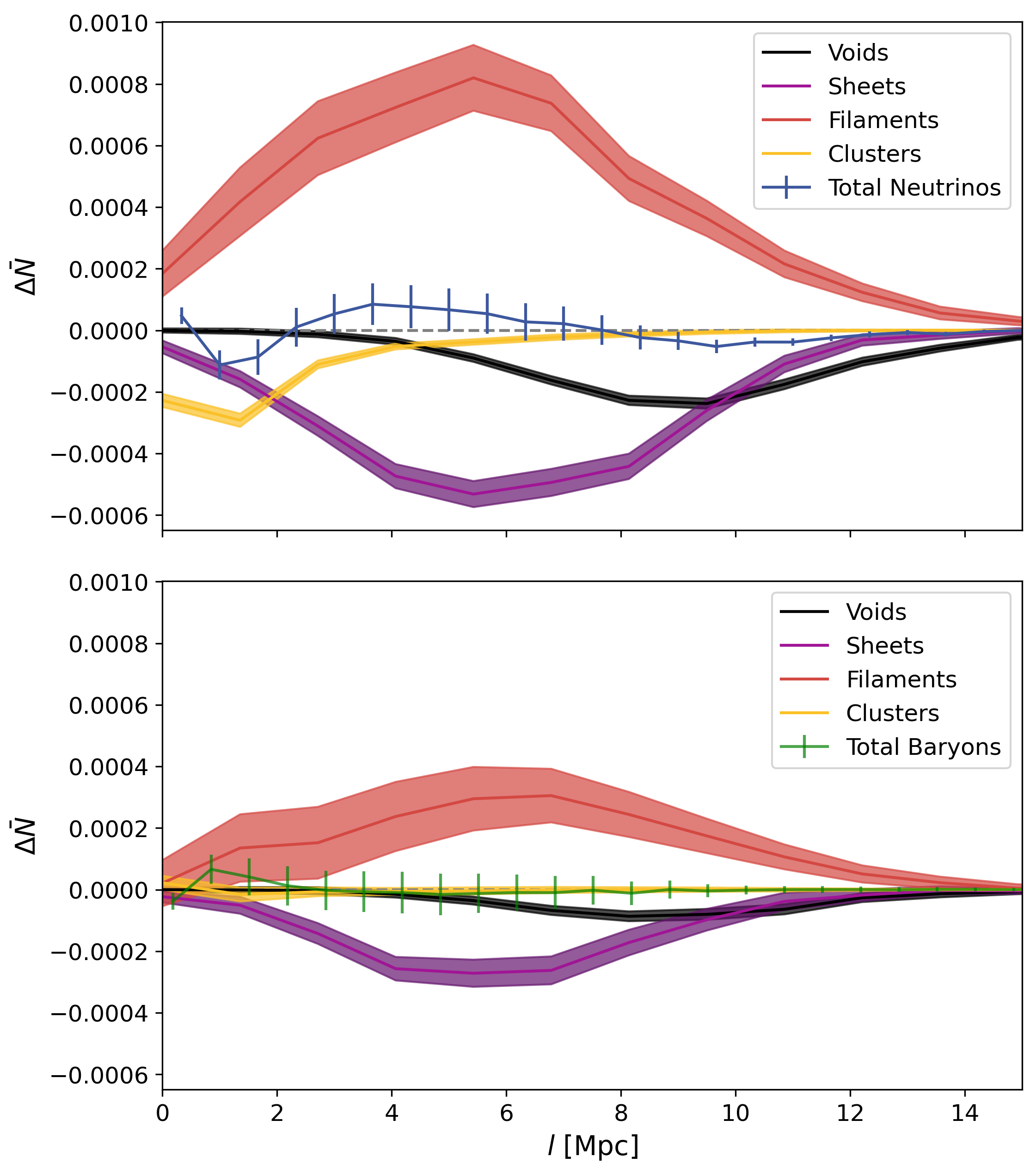}   
        \caption{The edge length statistic ($l$) split by each cosmic web environment, compared to the total statistic from Fig. \ref{fig:mst_comp} (labelled 'Total Neutrinos/Baryons'). In the top panel we plot the effect of massive neutrinos, subtracting the $M_\nu = 0.06\:{\rm eV\, }$ from the $M_\nu = 0.48\:{\rm eV\, }$ simulation, both DMO$+\nu$. In the bottom panel we plot the effect of baryons, subtract the DMO$+\nu$ from the HYDRO simulation, both with $M_\nu = 0.06\:{\rm eV\, }$.}
        \label{fig:mst_plit} 
\end{figure}

Motivated by the characteristic signals in edge length ($l$) seen in Fig. \ref{fig:mst_comp}, we now explore that same statistic but split by environment. We split the edges by environment as described in Section \ref{sec:edgeslie} and create figures similar to Fig. \ref{fig:edges_neutrinos}, where we split between environments, for the simulations: DMO $M_\nu=0.06 \:{\rm eV\, }$, DMO $M_\nu=0.48\:{\rm eV\, }$, and HYDRO $M_\nu=0.06\:{\rm eV\, }$. We then subtract the values from the DMO $M_\nu=0.06\:{\rm eV\, }$ simulation from the two mentioned previously to obtain the differences for each individual cosmic web component. The results are plotted in Fig. \ref{fig:mst_plit}, and we also overlay the total signal for direct comparison.

From these figures we link the edge length signal at different scales to the different cosmic web environments. Let us focus on the effect of massive neutrinos shown in the top panel of Fig. \ref{fig:mst_plit} since the baryonic signal (bottom) is noisy. At small scales of $l\approx 1\,\mathrm{Mpc}$, neutrinos decrease the number of edges which is associated with a decrease in edges in clusters and sheets at those scales. The increase at the scales of $l \approx 3\,\mathrm{Mpc}$ is linked to an increase of edges in filaments and the fact that cluster effects become less prominent at those scales. Finally, on larger scales of $l \approx 9\,\mathrm{Mpc}$ the decrease in number of edge lengths comes from the fact that at those scales voids become important and we also see a decrease of edges in voids. These MST results are capturing the overall effect of massive neutrinos that reshape the connectivity of the cosmic web across scales, suppressing dense environments while enhancing intermediate structures, and ultimately leading to more diffuse and less interconnected structures.

\section{Discussion}
In this paper, we explored the effects of massive neutrinos and baryons on the cosmic web using the FLAMINGO simulations. Using the NEXUS+ algorithm, we classified the cosmic web into voids, sheets, filaments, and clusters, and examined how their distribution is affected by the presence of massive neutrinos and the inclusion of  baryonic physics. Furthermore, we explore the relation between the Minimum Spanning Trees (MSTs), constructed from subhaloes, to the underlying cosmic web environment and its sensitivity to neutrino mass and baryons.

We find that increasing the neutrino mass shifts the density distribution towards higher densities and leads to  a  narrower density distribution, characterized by an increase of filamentary structures at the expense of voids and sheets. The fact that the width of the density distribution decreases can be interpreted as a reflection of  delayed structure formation in the presence of massive neutrinos: there is a smaller difference between densities in clusters and voids. This can be linked to the massive neutrinos slowing down structure formation since, as the Universe  ages, clusters become denser and voids become emptier.

Motivated by the MST's sensitivity to neutrino mass \citep{naidoo21neutrino}, we construct the MST from subhalo catalogues with varying neutrino mass. We show  the MST edges are predominantly located in filaments, confirming the MST's sensitivity to filamentary features of the cosmic web. Looking at the MST statistics, we find that the distribution of edge lengths is sensitive to the presence of massive neutrinos and baryonic physics in distinct ways. While neutrinos suppress small-scale ($l \leq 2\,\mathrm{Mpc}$) structure by smoothing filaments, baryons tend to increase the number of small-scale connections. Neutrinos then increase the number of edges at intermediate scales, $l \approx 4\,\mathrm{Mpc}$, which are the scales at which edges in filaments peak (see Fig. \ref{fig:edges_neutrinos}). We also see that massive neutrinos result in a decrease in the number of edges at longer lengths of $l \approx 9\,\mathrm{Mpc}$. The effects of massive neutrinos on the edge lengths at different scales suggest that the MST is sensitive to neutrino effects in different environments.

We linked the MST statistics to the effects on the density distribution described above and the difference in edge lengths split by environment. Massive neutrinos delay structure formation, leading to less dense clusters which is captured as a decrease in edge lengths at small scales. It also leads to less deep voids which is seen as a decrease in edge lengths at larger scales. The overall increase of edge lengths at intermediate scales is attributed to the increase of filaments and the fact that the environment is less developed.

It is worth noting that the neutrino masses used in this work are higher than the values we expect from the current constraints. The exaggerated effects can be used to study the consequences of massive neutrinos but we should explore further what is the sensitivity needed to pick up these effects in the case  of lower masses. Given the small current constraints on neutrino mass, it is possible that we are in a regime where, although baryonic effects are small, so too might be the neutrino effect, which will make it harder to distinguish the two: we emphasise the need for more statistics and more data to help break these degeneracies. However, given the tensions between DESI BAO and CMB data in current neutrino constraints, there is a strong motivation for alternative constraints from other large-scale structure probes. Other approaches like MST could help in this regard when combined with traditional 2PCFs.
We also note that to understand how these results translate to real data we should explore the connection between subhaloes and galaxies, the redshift evolution of this effect, the effect of redshift space distortions, and the effect of survey systematics \citep{Naidoo24}. In the future, it would be interesting to explore how the MST signal varies with different hydrodynamical models available in the FLAMINGO simulations, to see how it is affected by the feedback strength. It would also be useful to compare with the LS8 simulation, which varies $\sigma_8$ at fixed neutrino mass, to see whether the MST can help distinguish neutrino mass from a variation in $\sigma_8$ \citep{Villaescusa_2018}.

Nevertheless, our results demonstrate that MST statistics provide complementary information to two-point clustering measurements and offer a promising avenue for breaking the degeneracy between neutrino mass effects and baryon effects, leading to future improvements on neutrino mass constraints.  This work highlights the potential of using new cosmic web-sensitive probes, such as the MST, in upcoming large-scale structure surveys to improve constraints on cosmological parameters, in particular on neutrino mass.

\section*{Acknowledgements}
We thank Joop Schaye and Matthieu Schaller for their assistance with the data and for providing valuable feedback. We thank Ofer Lahav for providing insightful suggestions and comments. 

LNLS is supported by the Science and Technology Facilities Council (STFC) [grant number ST/Y509826/1]. KN acknowledges support from the Royal Society grant number URF\textbackslash R\textbackslash 231006. BJ acknowledges support by the ERC-selected UKRI Frontier Research Grant EP/Y03015X/1. WE and CSF acknowledge STFC Consolidated Grant ST/X001075/1 and support from the European Research Council (ERC) Advanced Investigator grant DMIDAS (GA 786910).

The simulations used in this project made use
of the DiRAC@Durham facility managed by the Institute for Computational Cosmology on behalf of the STFC DiRAC HPC Facility (\href{www.dirac.ac.uk}{www.dirac.ac.uk}). The equipment was funded by BEIS capital funding via STFC capital grants ST/K00042X/1, ST/P002293/1,
ST/R002371/1 and ST/S002502/1, Durham University and STFC
operations grant ST/R000832/1. DiRAC is part of the National eInfrastructure.

%%%%%%%%%%%%%%%%%%%%%%%%%%%%%%%%%%%%%%%%%%%%%%%%%%
\section*{Data Availability}

The derived data products produced in this paper can be shared on reasonable request to the corresponding author.



\bibliographystyle{mnras}
\bibliography{references} % if your bibtex file is called example.bib

@ARTICLE{Alpaslan2014,
       author = {{Alpaslan}, Mehmet and {Robotham}, Aaron S.~G. and {Driver}, Simon and {Norberg}, Peder and {Baldry}, Ivan and {Bauer}, Amanda E. and {Bland-Hawthorn}, Joss and {Brown}, Michael and {Cluver}, Michelle and {Colless}, Matthew and {Foster}, Caroline and {Hopkins}, Andrew and {Van Kampen}, Eelco and {Kelvin}, Lee and {Lara-Lopez}, Maritza A. and {Liske}, Jochen and {Lopez-Sanchez}, Angel R. and {Loveday}, Jon and {McNaught-Roberts}, Tamsyn and {Merson}, Alexander and {Pimbblet}, Kevin},
        title = "{Galaxy And Mass Assembly (GAMA): the large-scale structure of galaxies and comparison to mock universes}",
      journal = {\mnras},
         year = 2014,
        month = feb,
       volume = {438},
       number = {1},
        pages = {177-194},
          doi = {10.1093/mnras/stt2136},
archivePrefix = {arXiv},
       eprint = {1311.1211},
 primaryClass = {astro-ph.CO},
       adsurl = {https://ui.adsabs.harvard.edu/abs/2014MNRAS.438..177A}
}

@ARTICLE{Bhavsar1988,
       author = {{Bhavsar}, Suketu P. and {Ling}, E. Nigel},
        title = "{II. Large-Scale Distribution of Galaxies: Filamentary Structure and Visual Bias}",
      journal = {\pasp},
         year = 1988,
        month = nov,
       volume = {100},
        pages = {1314},
          doi = {10.1086/132325},
       adsurl = {https://ui.adsabs.harvard.edu/abs/1988PASP..100.1314B}
}

@article{cautun2013nexus,
  author = {{Cautun}, Marius and {van de Weygaert}, Rien and {Jones}, Bernard J.~T.},
        title = "{NEXUS: tracing the cosmic web connection}",
      journal = {\mnras},
         year = 2013,
        month = feb,
       volume = {429},
       number = {2},
        pages = {1286-1308},
          doi = {10.1093/mnras/sts416},
archivePrefix = {arXiv},
       eprint = {1209.2043},
 primaryClass = {astro-ph.CO},
       adsurl = {https://ui.adsabs.harvard.edu/abs/2013MNRAS.429.1286C}
}

@article{forero2009Tweb,
  author = {{Forero-Romero}, J.~E. and {Hoffman}, Y. and {Gottl{\"o}ber}, S. and {Klypin}, A. and {Yepes}, G.},
        title = "{A dynamical classification of the cosmic web}",
      journal = {\mnras},
         year = 2009,
        month = jul,
       volume = {396},
       number = {3},
        pages = {1815-1824},
          doi = {10.1111/j.1365-2966.2009.14885.x},
archivePrefix = {arXiv},
       eprint = {0809.4135},
 primaryClass = {astro-ph},
       adsurl = {https://ui.adsabs.harvard.edu/abs/2009MNRAS.396.1815F}
}

@ARTICLE{Schaller2024,
       author = {{Schaller}, Matthieu and {Borrow}, Josh and {Draper}, Peter W. and {Ivkovic}, Mladen and {McAlpine}, Stuart and {Vandenbroucke}, Bert and {Bah{\'e}}, Yannick and {Chaikin}, Evgenii and {Chalk}, Aidan B.~G. and {Chan}, Tsang Keung and {Correa}, Camila and {van Daalen}, Marcel and {Elbers}, Willem and {Gonnet}, Pedro and {Hausammann}, Lo{\"\i}c and {Helly}, John and {Hu{\v{s}}ko}, Filip and {Kegerreis}, Jacob A. and {Nobels}, Folkert S.~J. and {Ploeckinger}, Sylvia and {Revaz}, Yves and {Roper}, William J. and {Ruiz-Bonilla}, Sergio and {Sandnes}, Thomas D. and {Uyttenhove}, Yolan and {Willis}, James S. and {Xiang}, Zhen},
        title = "{SWIFT: A modern highly-parallel gravity and smoothed particle hydrodynamics solver for astrophysical and cosmological applications}",
        journal = {\mnras},
         year = {2024},
        month = may,
       volume = {530},
       number = {2},
        pages = {2378-2419},
          doi = {10.1093/mnras/stae922},
archivePrefix = {arXiv},
       eprint = {2305.13380},
 primaryClass = {astro-ph.IM},
       adsurl ={https://ui.adsabs.harvard.edu/abs/2024MNRAS.530.2378S}
}

@ARTICLE{Elbers2021,
       author = {{Elbers}, Willem and {Frenk}, Carlos S. and {Jenkins}, Adrian and {Li}, Baojiu and {Pascoli}, Silvia},
        title = "{An optimal non-linear method for simulating relic neutrinos}",
      journal = {\mnras},
         year = 2021,
        month = oct,
       volume = {507},
       number = {2},
        pages = {2614-2631},
          doi = {10.1093/mnras/stab2260},
archivePrefix = {arXiv},
       eprint = {2010.07321},
 primaryClass = {astro-ph.CO},
       adsurl = {https://ui.adsabs.harvard.edu/abs/2021MNRAS.507.2614E}
}

@article{Naidoo2019Mistree,
doi = {10.21105/joss.01721}, url = {https://doi.org/10.21105/joss.01721}, year = {2019}, publisher = {The Open Journal}, volume = {4}, pages = {1721}, author = {Krishna Naidoo}, title = {MiSTree: a Python package for constructing and analysing Minimum Spanning Trees}, journal = {Journal of Open Source Software} }

@article{macdonald2005minimum,
  author = {{Macdonald}, P.~J. and {Almaas}, E. and {Barab{\'a}si}, A.-L.},
        title = "{Minimum spanning trees of weighted scale-free networks}",
      journal = {EPL (Europhysics Letters)},
         year = 2005,
        month = oct,
       volume = {72},
       number = {2},
        pages = {308-314},
          doi = {10.1209/epl/i2005-10232-x},
       adsurl = {https://ui.adsabs.harvard.edu/abs/2005EL.....72..308M}
}

@article{chang2011classification,
  author = {Chang, Y.F. and Chen, C.-M.},
title = {Classification and visualization of the social science network by the minimum span clustering method},
journal = {Journal of the American Society for Information Science and Technology},
volume = {62},
number = {12},
pages = {2404-2413},
doi = {https://doi.org/10.1002/asi.21634},
url = {https://onlinelibrary.wiley.com/doi/abs/10.1002/asi.21634},
eprint = {https://onlinelibrary.wiley.com/doi/pdf/10.1002/asi.21634},
year = {2011}
}

@article{chin1978CS_MST,
  title = {Algorithms for updating minimal spanning trees},
journal = {Journal of Computer and System Sciences},
volume = {16},
number = {3},
pages = {333-344},
year = {1978},
issn = {0022-0000},
doi = {https://doi.org/10.1016/0022-0000(78)90022-3},
url = {https://www.sciencedirect.com/science/article/pii/0022000078900223},
author = {Francis Chin and David Houck}
}

@article{barrow1985minimal,
  author = {{Barrow}, J.~D. and {Bhavsar}, S.~P. and {Sonoda}, D.~H.},
        title = "{Minimal spanning trees, filaments and galaxy clustering}",
      journal = {\mnras},
         year = 1985,
        month = sep,
       volume = {216},
        pages = {17-35},
          doi = {10.1093/mnras/216.1.17},
       adsurl = {https://ui.adsabs.harvard.edu/abs/1985MNRAS.216...17B}
}

@article{graham1995quasarMST,
  author = {{Graham}, M.~J. and {Clowes}, R.~G. and {Campusano}, L.~E.},
        title = "{Finding Quasar Superstructures}",
      journal = {\mnras},
         year = 1995,
        month = aug,
       volume = {275},
        pages = {790},
          doi = {10.1093/mnras/275.3.790},
       adsurl = {https://ui.adsabs.harvard.edu/abs/1995MNRAS.275..790G}
}

@article{bhavsar1988filamentsreal,
  author = {{Bhavsar}, Suketu P. and {Ling}, E. Nigel},
        title = "{Are the Filaments Real?}",
      journal = {\apjl},
         year = 1988,
        month = aug,
       volume = {331},
        pages = {L63},
          doi = {10.1086/185236},
       adsurl = {https://ui.adsabs.harvard.edu/abs/1988ApJ...331L..63B}
}

@article{colberg2007quantifying,
  author = {{Colberg}, J{\"o}rg M.},
        title = "{Quantifying cosmic superstructures}",
      journal = {\mnras},
         year = 2007,
        month = feb,
       volume = {375},
       number = {1},
        pages = {337-347},
          doi = {10.1111/j.1365-2966.2006.11312.x},
archivePrefix = {arXiv},
       eprint = {astro-ph/0611641},
 primaryClass = {astro-ph},
       adsurl = {https://ui.adsabs.harvard.edu/abs/2007MNRAS.375..337C}
}

@article{bond1996filaments,
  author = {{Bond}, J. Richard and {Kofman}, Lev and {Pogosyan}, Dmitry},
        title = "{How filaments of galaxies are woven into the cosmic web}",
      journal = {\nat},
         year = 1996,
        month = apr,
       volume = {380},
       number = {6575},
        pages = {603-606},
          doi = {10.1038/380603a0},
archivePrefix = {arXiv},
       eprint = {astro-ph/9512141},
 primaryClass = {astro-ph},
       adsurl = {https://ui.adsabs.harvard.edu/abs/1996Natur.380..603B}
}

@article{libeskind2018tracing,
  author = {{Libeskind}, Noam I. and {van de Weygaert}, Rien and {Cautun}, Marius and {Falck}, Bridget and {Tempel}, Elmo and {Abel}, Tom and {Alpaslan}, Mehmet and {Arag{\'o}n-Calvo}, Miguel A. and {Forero-Romero}, Jaime E. and {Gonzalez}, Roberto and {Gottl{\"o}ber}, Stefan and {Hahn}, Oliver and {Hellwing}, Wojciech A. and {Hoffman}, Yehuda and {Jones}, Bernard J.~T. and {Kitaura}, Francisco and {Knebe}, Alexander and {Manti}, Serena and {Neyrinck}, Mark and {Nuza}, Sebasti{\'a}n E. and {Padilla}, Nelson and {Platen}, Erwin and {Ramachandra}, Nesar and {Robotham}, Aaron and {Saar}, Enn and {Shandarin}, Sergei and {Steinmetz}, Matthias and {Stoica}, Radu S. and {Sousbie}, Thierry and {Yepes}, Gustavo},
        title = "{Tracing the cosmic web}",
      journal = {\mnras},
         year = 2018,
        month = jan,
       volume = {473},
       number = {1},
        pages = {1195-1217},
          doi = {10.1093/mnras/stx1976},
archivePrefix = {arXiv},
       eprint = {1705.03021},
 primaryClass = {astro-ph.CO},
       adsurl = {https://ui.adsabs.harvard.edu/abs/2018MNRAS.473.1195L}
}

@ARTICLE{Zeldovich1970,
       author = {{Zel'dovich}, Ya. B.},
        title = "{Gravitational instability: An approximate theory for large density perturbations.}",
      journal = {\aap},
         year = 1970,
        month = mar,
       volume = {5},
        pages = {84-89},
       adsurl = {https://ui.adsabs.harvard.edu/abs/1970A&A.....5...84Z}
}

@article{collaboration2020planck,
  author = {{Planck Collaboration} and {Aghanim}, N. and {Akrami}, Y. and {Ashdown}, M. and {Aumont}, J. and {Baccigalupi}, C. and {Ballardini}, M. and {Banday}, A.~J. and {Barreiro}, R.~B. and {Bartolo}, N. and {Basak}, S. and {Battye}, R. and {Benabed}, K. and {Bernard}, J.-P. and {Bersanelli}, M. and {Bielewicz}, P. and {Bock}, J.~J. and {Bond}, J.~R. and {Borrill}, J. and {Bouchet}, F.~R. and {Boulanger}, F. and {Bucher}, M. and {Burigana}, C. and {Butler}, R.~C. and {Calabrese}, E. and {Cardoso}, J.-F. and {Carron}, J. and {Challinor}, A. and {Chiang}, H.~C. and {Chluba}, J. and {Colombo}, L.~P.~L. and {Combet}, C. and {Contreras}, D. and {Crill}, B.~P. and {Cuttaia}, F. and {de Bernardis}, P. and {de Zotti}, G. and {Delabrouille}, J. and {Delouis}, J.-M. and {Di Valentino}, E. and {Diego}, J.~M. and {Dor{\'e}}, O. and {Douspis}, M. and {Ducout}, A. and {Dupac}, X. and {Dusini}, S. and {Efstathiou}, G. and {Elsner}, F. and {En{\ss}lin}, T.~A. and {Eriksen}, H.~K. and {Fantaye}, Y. and {Farhang}, M. and {Fergusson}, J. and {Fernandez-Cobos}, R. and {Finelli}, F. and {Forastieri}, F. and {Frailis}, M. and {Fraisse}, A.~A. and {Franceschi}, E. and {Frolov}, A. and {Galeotta}, S. and {Galli}, S. and {Ganga}, K. and {G{\'e}nova-Santos}, R.~T. and {Gerbino}, M. and {Ghosh}, T. and {Gonz{\'a}lez-Nuevo}, J. and {G{\'o}rski}, K.~M. and {Gratton}, S. and {Gruppuso}, A. and {Gudmundsson}, J.~E. and {Hamann}, J. and {Handley}, W. and {Hansen}, F.~K. and {Herranz}, D. and {Hildebrandt}, S.~R. and {Hivon}, E. and {Huang}, Z. and {Jaffe}, A.~H. and {Jones}, W.~C. and {Karakci}, A. and {Keih{\"a}nen}, E. and {Keskitalo}, R. and {Kiiveri}, K. and {Kim}, J. and {Kisner}, T.~S. and {Knox}, L. and {Krachmalnicoff}, N. and {Kunz}, M. and {Kurki-Suonio}, H. and {Lagache}, G. and {Lamarre}, J.-M. and {Lasenby}, A. and {Lattanzi}, M. and {Lawrence}, C.~R. and {Le Jeune}, M. and {Lemos}, P. and {Lesgourgues}, J. and {Levrier}, F. and {Lewis}, A. and {Liguori}, M. and {Lilje}, P.~B. and {Lilley}, M. and {Lindholm}, V. and {L{\'o}pez-Caniego}, M. and {Lubin}, P.~M. and {Ma}, Y.-Z. and {Mac{\'\i}as-P{\'e}rez}, J.~F. and {Maggio}, G. and {Maino}, D. and {Mandolesi}, N. and {Mangilli}, A. and {Marcos-Caballero}, A. and {Maris}, M. and {Martin}, P.~G. and {Martinelli}, M. and {Mart{\'\i}nez-Gonz{\'a}lez}, E. and {Matarrese}, S. and {Mauri}, N. and {McEwen}, J.~D. and {Meinhold}, P.~R. and {Melchiorri}, A. and {Mennella}, A. and {Migliaccio}, M. and {Millea}, M. and {Mitra}, S. and {Miville-Desch{\^e}nes}, M.-A. and {Molinari}, D. and {Montier}, L. and {Morgante}, G. and {Moss}, A. and {Natoli}, P. and {N{\o}rgaard-Nielsen}, H.~U. and {Pagano}, L. and {Paoletti}, D. and {Partridge}, B. and {Patanchon}, G. and {Peiris}, H.~V. and {Perrotta}, F. and {Pettorino}, V. and {Piacentini}, F. and {Polastri}, L. and {Polenta}, G. and {Puget}, J.-L. and {Rachen}, J.~P. and {Reinecke}, M. and {Remazeilles}, M. and {Renzi}, A. and {Rocha}, G. and {Rosset}, C. and {Roudier}, G. and {Rubi{\~n}o-Mart{\'\i}n}, J.~A. and {Ruiz-Granados}, B. and {Salvati}, L. and {Sandri}, M. and {Savelainen}, M. and {Scott}, D. and {Shellard}, E.~P.~S. and {Sirignano}, C. and {Sirri}, G. and {Spencer}, L.~D. and {Sunyaev}, R. and {Suur-Uski}, A.-S. and {Tauber}, J.~A. and {Tavagnacco}, D. and {Tenti}, M. and {Toffolatti}, L. and {Tomasi}, M. and {Trombetti}, T. and {Valenziano}, L. and {Valiviita}, J. and {Van Tent}, B. and {Vibert}, L. and {Vielva}, P. and {Villa}, F. and {Vittorio}, N. and {Wandelt}, B.~D. and {Wehus}, I.~K. and {White}, M. and {White}, S.~D.~M. and {Zacchei}, A. and {Zonca}, A.},
        title = "{Planck 2018 results. VI. Cosmological parameters}",
      journal = {\aap},
         year = 2020,
        month = sep,
       volume = {641},
          eid = {A6},
        pages = {A6},
          doi = {10.1051/0004-6361/201833910},
archivePrefix = {arXiv},
       eprint = {1807.06209},
 primaryClass = {astro-ph.CO},
       adsurl = {https://ui.adsabs.harvard.edu/abs/2020A&A...641A...6P}
}

@article{naidoo21neutrino,
    author = {{Naidoo}, Krishna and {Massara}, Elena and {Lahav}, Ofer},
        title = "{Cosmology and neutrino mass with the minimum spanning tree}",
      journal = {\mnras},
         year = 2022,
        month = jul,
       volume = {513},
       number = {3},
        pages = {3596-3609},
          doi = {10.1093/mnras/stac1138},
archivePrefix = {arXiv},
       eprint = {2111.12088},
 primaryClass = {astro-ph.CO},
       adsurl = {https://ui.adsabs.harvard.edu/abs/2022MNRAS.513.3596N}
}

@article{alam2017clustering,
   author = {{Alam}, Shadab and {Ata}, Metin and {Bailey}, Stephen and {Beutler}, Florian and {Bizyaev}, Dmitry and {Blazek}, Jonathan A. and {Bolton}, Adam S. and {Brownstein}, Joel R. and {Burden}, Angela and {Chuang}, Chia-Hsun and {Comparat}, Johan and {Cuesta}, Antonio J. and {Dawson}, Kyle S. and {Eisenstein}, Daniel J. and {Escoffier}, Stephanie and {Gil-Mar{\'\i}n}, H{\'e}ctor and {Grieb}, Jan Niklas and {Hand}, Nick and {Ho}, Shirley and {Kinemuchi}, Karen and {Kirkby}, David and {Kitaura}, Francisco and {Malanushenko}, Elena and {Malanushenko}, Viktor and {Maraston}, Claudia and {McBride}, Cameron K. and {Nichol}, Robert C. and {Olmstead}, Matthew D. and {Oravetz}, Daniel and {Padmanabhan}, Nikhil and {Palanque-Delabrouille}, Nathalie and {Pan}, Kaike and {Pellejero-Ibanez}, Marcos and {Percival}, Will J. and {Petitjean}, Patrick and {Prada}, Francisco and {Price-Whelan}, Adrian M. and {Reid}, Beth A. and {Rodr{\'\i}guez-Torres}, Sergio A. and {Roe}, Natalie A. and {Ross}, Ashley J. and {Ross}, Nicholas P. and {Rossi}, Graziano and {Rubi{\~n}o-Mart{\'\i}n}, Jose Alberto and {Saito}, Shun and {Salazar-Albornoz}, Salvador and {Samushia}, Lado and {S{\'a}nchez}, Ariel G. and {Satpathy}, Siddharth and {Schlegel}, David J. and {Schneider}, Donald P. and {Sc{\'o}ccola}, Claudia G. and {Seo}, Hee-Jong and {Sheldon}, Erin S. and {Simmons}, Audrey and {Slosar}, An{\v{z}}e and {Strauss}, Michael A. and {Swanson}, Molly E.~C. and {Thomas}, Daniel and {Tinker}, Jeremy L. and {Tojeiro}, Rita and {Maga{\~n}a}, Mariana Vargas and {Vazquez}, Jose Alberto and {Verde}, Licia and {Wake}, David A. and {Wang}, Yuting and {Weinberg}, David H. and {White}, Martin and {Wood-Vasey}, W. Michael and {Y{\`e}che}, Christophe and {Zehavi}, Idit and {Zhai}, Zhongxu and {Zhao}, Gong-Bo},
        title = "{The clustering of galaxies in the completed SDSS-III Baryon Oscillation Spectroscopic Survey: cosmological analysis of the DR12 galaxy sample}",
      journal = {\mnras},
         year = 2017,
        month = sep,
       volume = {470},
       number = {3},
        pages = {2617-2652},
          doi = {10.1093/mnras/stx721},
archivePrefix = {arXiv},
       eprint = {1607.03155},
 primaryClass = {astro-ph.CO},
       adsurl = {https://ui.adsabs.harvard.edu/abs/2017MNRAS.470.2617A}
}

@article{abbott2018WL,
  author = {{Abbott}, T.~M.~C. and {Abdalla}, F.~B. and {Alarcon}, A. and {Aleksi{\'c}}, J. and {Allam}, S. and {Allen}, S. and {Amara}, A. and {Annis}, J. and {Asorey}, J. and {Avila}, S. and {Bacon}, D. and {Balbinot}, E. and {Banerji}, M. and {Banik}, N. and {Barkhouse}, W. and {Baumer}, M. and {Baxter}, E. and {Bechtol}, K. and {Becker}, M.~R. and {Benoit-L{\'e}vy}, A. and {Benson}, B.~A. and {Bernstein}, G.~M. and {Bertin}, E. and {Blazek}, J. and {Bridle}, S.~L. and {Brooks}, D. and {Brout}, D. and {Buckley-Geer}, E. and {Burke}, D.~L. and {Busha}, M.~T. and {Campos}, A. and {Capozzi}, D. and {Carnero Rosell}, A. and {Carrasco Kind}, M. and {Carretero}, J. and {Castander}, F.~J. and {Cawthon}, R. and {Chang}, C. and {Chen}, N. and {Childress}, M. and {Choi}, A. and {Conselice}, C. and {Crittenden}, R. and {Crocce}, M. and {Cunha}, C.~E. and {D'Andrea}, C.~B. and {da Costa}, L.~N. and {Das}, R. and {Davis}, T.~M. and {Davis}, C. and {De Vicente}, J. and {DePoy}, D.~L. and {DeRose}, J. and {Desai}, S. and {Diehl}, H.~T. and {Dietrich}, J.~P. and {Dodelson}, S. and {Doel}, P. and {Drlica-Wagner}, A. and {Eifler}, T.~F. and {Elliott}, A.~E. and {Elsner}, F. and {Elvin-Poole}, J. and {Estrada}, J. and {Evrard}, A.~E. and {Fang}, Y. and {Fernandez}, E. and {Fert{\'e}}, A. and {Finley}, D.~A. and {Flaugher}, B. and {Fosalba}, P. and {Friedrich}, O. and {Frieman}, J. and {Garc{\'\i}a-Bellido}, J. and {Garcia-Fernandez}, M. and {Gatti}, M. and {Gaztanaga}, E. and {Gerdes}, D.~W. and {Giannantonio}, T. and {Gill}, M.~S.~S. and {Glazebrook}, K. and {Goldstein}, D.~A. and {Gruen}, D. and {Gruendl}, R.~A. and {Gschwend}, J. and {Gutierrez}, G. and {Hamilton}, S. and {Hartley}, W.~G. and {Hinton}, S.~R. and {Honscheid}, K. and {Hoyle}, B. and {Huterer}, D. and {Jain}, B. and {James}, D.~J. and {Jarvis}, M. and {Jeltema}, T. and {Johnson}, M.~D. and {Johnson}, M.~W.~G. and {Kacprzak}, T. and {Kent}, S. and {Kim}, A.~G. and {King}, A. and {Kirk}, D. and {Kokron}, N. and {Kovacs}, A. and {Krause}, E. and {Krawiec}, C. and {Kremin}, A. and {Kuehn}, K. and {Kuhlmann}, S. and {Kuropatkin}, N. and {Lacasa}, F. and {Lahav}, O. and {Li}, T.~S. and {Liddle}, A.~R. and {Lidman}, C. and {Lima}, M. and {Lin}, H. and {MacCrann}, N. and {Maia}, M.~A.~G. and {Makler}, M. and {Manera}, M. and {March}, M. and {Marshall}, J.~L. and {Martini}, P. and {McMahon}, R.~G. and {Melchior}, P. and {Menanteau}, F. and {Miquel}, R. and {Miranda}, V. and {Mudd}, D. and {Muir}, J. and {M{\"o}ller}, A. and {Neilsen}, E. and {Nichol}, R.~C. and {Nord}, B. and {Nugent}, P. and {Ogando}, R.~L.~C. and {Palmese}, A. and {Peacock}, J. and {Peiris}, H.~V. and {Peoples}, J. and {Percival}, W.~J. and {Petravick}, D. and {Plazas}, A.~A. and {Porredon}, A. and {Prat}, J. and {Pujol}, A. and {Rau}, M.~M. and {Refregier}, A. and {Ricker}, P.~M. and {Roe}, N. and {Rollins}, R.~P. and {Romer}, A.~K. and {Roodman}, A. and {Rosenfeld}, R. and {Ross}, A.~J. and {Rozo}, E. and {Rykoff}, E.~S. and {Sako}, M. and {Salvador}, A.~I. and {Samuroff}, S. and {S{\'a}nchez}, C. and {Sanchez}, E. and {Santiago}, B. and {Scarpine}, V. and {Schindler}, R. and {Scolnic}, D. and {Secco}, L.~F. and {Serrano}, S. and {Sevilla-Noarbe}, I. and {Sheldon}, E. and {Smith}, R.~C. and {Smith}, M. and {Smith}, J. and {Soares-Santos}, M. and {Sobreira}, F. and {Suchyta}, E. and {Tarle}, G. and {Thomas}, D. and {Troxel}, M.~A. and {Tucker}, D.~L. and {Tucker}, B.~E. and {Uddin}, S.~A. and {Varga}, T.~N. and {Vielzeuf}, P. and {Vikram}, V. and {Vivas}, A.~K. and {Walker}, A.~R. and {Wang}, M. and {Wechsler}, R.~H. and {Weller}, J. and {Wester}, W. and {Wolf}, R.~C. and {Yanny}, B. and {Yuan}, F. and {Zenteno}, A. and {Zhang}, B. and {Zhang}, Y. and {Zuntz}, J.},
        title = "{Dark Energy Survey year 1 results: Cosmological constraints from galaxy clustering and weak lensing}",
      journal = {\prd},
         year = 2018,
        month = aug,
       volume = {98},
       number = {4},
          eid = {043526},
        pages = {043526},
          doi = {10.1103/PhysRevD.98.043526},
archivePrefix = {arXiv},
       eprint = {1708.01530},
 primaryClass = {astro-ph.CO},
       adsurl = {https://ui.adsabs.harvard.edu/abs/2018PhRvD..98d3526A}
}

@article{de2019baryon,
  author = {{de Sainte Agathe}, Victoria and {Balland}, Christophe and {du Mas des Bourboux}, H{\'e}lion and {Busca}, Nicol{\'a}s G. and {Blomqvist}, Michael and {Guy}, Julien and {Rich}, James and {Font-Ribera}, Andreu and {Pieri}, Matthew M. and {Bautista}, Julian E. and {Dawson}, Kyle and {Le Goff}, Jean-Marc and {de la Macorra}, Axel and {Palanque-Delabrouille}, Nathalie and {Percival}, Will J. and {P{\'e}rez-R{\`a}fols}, Ignasi and {Schneider}, Donald P. and {Slosar}, An{\v{z}}e and {Y{\`e}che}, Christophe},
        title = "{Baryon acoustic oscillations at z = 2.34 from the correlations of Ly{\ensuremath{\alpha}} absorption in eBOSS DR14}",
      journal = {\aap},
         year = 2019,
        month = sep,
       volume = {629},
          eid = {A85},
        pages = {A85},
          doi = {10.1051/0004-6361/201935638},
archivePrefix = {arXiv},
       eprint = {1904.03400},
 primaryClass = {astro-ph.CO},
       adsurl = {https://ui.adsabs.harvard.edu/abs/2019A&A...629A..85D}
}

@article{riess2016candles,
  author = {{Riess}, Adam G. and {Macri}, Lucas M. and {Hoffmann}, Samantha L. and {Scolnic}, Dan and {Casertano}, Stefano and {Filippenko}, Alexei V. and {Tucker}, Brad E. and {Reid}, Mark J. and {Jones}, David O. and {Silverman}, Jeffrey M. and {Chornock}, Ryan and {Challis}, Peter and {Yuan}, Wenlong and {Brown}, Peter J. and {Foley}, Ryan J.},
        title = "{A 2.4\% Determination of the Local Value of the Hubble Constant}",
      journal = {\apj},
         year = 2016,
        month = jul,
       volume = {826},
       number = {1},
          eid = {56},
        pages = {56},
          doi = {10.3847/0004-637X/826/1/56},
archivePrefix = {arXiv},
       eprint = {1604.01424},
 primaryClass = {astro-ph.CO},
       adsurl = {https://ui.adsabs.harvard.edu/abs/2016ApJ...826...56R}
}

@article{slepian20173PFC,
  author = {{Slepian}, Zachary and {Eisenstein}, Daniel J. and {Beutler}, Florian and {Chuang}, Chia-Hsun and {Cuesta}, Antonio J. and {Ge}, Jian and {Gil-Mar{\'\i}n}, H{\'e}ctor and {Ho}, Shirley and {Kitaura}, Francisco-Shu and {McBride}, Cameron K. and {Nichol}, Robert C. and {Percival}, Will J. and {Rodr{\'\i}guez-Torres}, Sergio and {Ross}, Ashley J. and {Scoccimarro}, Rom{\'a}n and {Seo}, Hee-Jong and {Tinker}, Jeremy and {Tojeiro}, Rita and {Vargas-Maga{\~n}a}, Mariana},
        title = "{The large-scale three-point correlation function of the SDSS BOSS DR12 CMASS galaxies}",
      journal = {\mnras},
         year = 2017,
        month = jun,
       volume = {468},
       number = {1},
        pages = {1070-1083},
          doi = {10.1093/mnras/stw3234},
archivePrefix = {arXiv},
       eprint = {1512.02231},
 primaryClass = {astro-ph.CO},
       adsurl = {https://ui.adsabs.harvard.edu/abs/2017MNRAS.468.1070S}
}

@article{gil2016bspectrum,
  author = {{Gil-Mar{\'\i}n}, H{\'e}ctor and {Percival}, Will J. and {Verde}, Licia and {Brownstein}, Joel R. and {Chuang}, Chia-Hsun and {Kitaura}, Francisco-Shu and {Rodr{\'\i}guez-Torres}, Sergio A. and {Olmstead}, Matthew D.},
        title = "{The clustering of galaxies in the SDSS-III Baryon Oscillation Spectroscopic Survey: RSD measurement from the power spectrum and bispectrum of the DR12 BOSS galaxies}",
      journal = {\mnras},
         year = 2017,
        month = feb,
       volume = {465},
       number = {2},
        pages = {1757-1788},
          doi = {10.1093/mnras/stw2679},
archivePrefix = {arXiv},
       eprint = {1606.00439},
 primaryClass = {astro-ph.CO},
       adsurl = {https://ui.adsabs.harvard.edu/abs/2017MNRAS.465.1757G}
}

@article{kruskal1956shortest,
  ISSN = {00029939, 10886826},
  URL = {http://www.jstor.org/stable/2033241},
  author = {Kruskal, Joseph B.},
  title = {On the Shortest Spanning Subtree of a Graph and the Traveling Salesman Problem},
  journal = {\href{http://www.jstor.org/stable/2033241}{Proceedings of the American Mathematical Society, 7, 48}},
  number = {1},
  publisher = {American Mathematical Society},
  year = {1956},
  urldate = {2025-12-09}
}

@article{rainbolt2017use,
  author = {{Rainbolt}, J. Lovelace and {Schmitt}, M.},
        title = "{The use of minimal spanning trees in particle physics}",
      journal = {Journal of Instrumentation},
         year = 2017,
        month = feb,
       volume = {12},
       number = {2},
        pages = {P02009},
          doi = {10.1088/1748-0221/12/02/P02009},
archivePrefix = {arXiv},
       eprint = {1608.04772},
 primaryClass = {stat.AP},
       adsurl = {https://ui.adsabs.harvard.edu/abs/2017JInst..12P2009L}
}

@ARTICLE{Schaap2000,
       author = {{Schaap}, W.~E. and {van de Weygaert}, R.},
        title = "{Continuous fields and discrete samples: reconstruction through Delaunay tessellations}",
      journal = {\aap},
         year = 2000,
        month = nov,
       volume = {363},
        pages = {L29-L32},
          doi = {10.48550/arXiv.astro-ph/0011007},
archivePrefix = {arXiv},
       eprint = {astro-ph/0011007},
 primaryClass = {astro-ph},
       adsurl = {https://ui.adsabs.harvard.edu/abs/2000A&A...363L..29S}
}

@ARTICLE{MMF2010,
       author = {{Arag{\'o}n-Calvo}, Miguel A. and {van de Weygaert}, Rien and {Jones}, Bernard J.~T.},
        title = "{Multiscale phenomenology of the cosmic web}",
      journal = {\mnras},
         year = 2010,
        month = nov,
       volume = {408},
       number = {4},
        pages = {2163-2187},
          doi = {10.1111/j.1365-2966.2010.17263.x},
archivePrefix = {arXiv},
       eprint = {1007.0742},
 primaryClass = {astro-ph.CO},
       adsurl = {https://ui.adsabs.harvard.edu/abs/2010MNRAS.408.2163A}
}

@ARTICLE{Hoffman2012,
       author = {{Hoffman}, Yehuda and {Metuki}, Ofer and {Yepes}, Gustavo and {Gottl{\"o}ber}, Stefan and {Forero-Romero}, Jaime E. and {Libeskind}, Noam I. and {Knebe}, Alexander},
        title = "{A kinematic classification of the cosmic web}",
      journal = {\mnras},
         year = 2012,
        month = sep,
       volume = {425},
       number = {3},
        pages = {2049-2057},
          doi = {10.1111/j.1365-2966.2012.21553.x},
archivePrefix = {arXiv},
       eprint = {1201.3367},
 primaryClass = {astro-ph.CO},
       adsurl = {https://ui.adsabs.harvard.edu/abs/2012MNRAS.425.2049H}
}

@article{naidoo2020beyond,
  author = {{Naidoo}, Krishna and {Whiteway}, Lorne and {Massara}, Elena and {Gualdi}, Davide and {Lahav}, Ofer and {Viel}, Matteo and {Gil-Mar{\'\i}n}, H{\'e}ctor and {Font-Ribera}, Andreu},
        title = "{Beyond two-point statistics: using the minimum spanning tree as a tool for cosmology}",
      journal = {\mnras},
         year = 2020,
        month = jan,
       volume = {491},
       number = {2},
        pages = {1709-1726},
          doi = {10.1093/mnras/stz3075},
archivePrefix = {arXiv},
       eprint = {1907.00989},
 primaryClass = {astro-ph.CO},
       adsurl = {https://ui.adsabs.harvard.edu/abs/2020MNRAS.491.1709N}
}

@article{miller1974jackknife,
  ISSN = {00063444, 14643510},
 URL ={http://www.jstor.org/stable/2334280},
 author = {Rupert G. Miller},
 journal = {Biometrika},
 number = {1},
 pages = {1--15},
 publisher = {[Oxford University Press, Biometrika Trust]},
 title = {The Jackknife--A Review},
 urldate = {2025-12-09},
 volume = {61},
 year = {1974}
}

@article{agarwal2011neutPP,
       author = {{Agarwal}, Shankar and {Feldman}, Hume A.},
        title = "{The effect of massive neutrinos on the matter power spectrum}",
      journal = {\mnras},
         year = 2011,
        month = jan,
       volume = {410},
       number = {3},
        pages = {1647-1654},
          doi = {10.1111/j.1365-2966.2010.17546.x},
archivePrefix = {arXiv},
       eprint = {1006.0689},
 primaryClass = {astro-ph.CO},
       adsurl = {https://ui.adsabs.harvard.edu/abs/2011MNRAS.410.1647A}
}

@article{schneider2019BarPP,
  author = {{Schneider}, Aurel and {Teyssier}, Romain and {Stadel}, Joachim and {Chisari}, Nora Elisa and {Le Brun}, Amandine M.~C. and {Amara}, Adam and {Refregier}, Alexandre},
        title = "{Quantifying baryon effects on the matter power spectrum and the weak lensing shear correlation}",
      journal = {\jcap},
         year = 2019,
        month = mar,
       volume = {2019},
       number = {3},
          eid = {020},
        pages = {020},
          doi = {10.1088/1475-7516/2019/03/020},
archivePrefix = {arXiv},
       eprint = {1810.08629},
 primaryClass = {astro-ph.CO},
       adsurl = {https://ui.adsabs.harvard.edu/abs/2019JCAP...03..020S}
}

@ARTICLE{galaxy_density23,
       author = {{Prada}, F. and {Ereza}, J. and {Smith}, A. and {Lasker}, J. and {Vaisakh}, R. and {Kehoe}, R. and {Dong-P{\'a}ez}, C.~A. and {Siudek}, M. and {Wang}, M.~S. and {Alam}, S. and {Beutler}, F. and {Bianchi}, D. and {Cole}, S. and {Dey}, B. and {Kirkby}, D. and {Norberg}, P. and {Aguilar}, J. and {Ahlen}, S. and {Brooks}, D. and {Claybaugh}, T. and {Dawson}, K. and {de la Macorra}, A. and {Fanning}, K. and {Forero-Romero}, J.~E. and {Gontcho A Gontcho}, S. and {Hahn}, C. and {Honscheid}, K. and {Ishak}, M. and {Kisner}, T. and {Landriau}, M. and {Manera}, M. and {Meisner}, A. and {Miquel}, R. and {Moustakas}, J. and {Mueller}, E. and {Nie}, J. and {Percival}, W.~J. and {Poppett}, C. and {Rezaie}, M. and {Rossi}, G. and {Sanchez}, E. and {Schubnell}, M. and {Tarl{\'e}}, G. and {Vargas-Maga{\~n}a}, M. and {Weaver}, B.~A. and {Yuan}, S. and {Zhou}, Z.},
        title = "{The DESI One-Percent Survey: Modelling the clustering and halo occupation of all four DESI tracers with UCHUU}",
      journal = {\aap},
         year = 2025,
        month = jun,
       volume = {698},
          eid = {A170},
        pages = {A170},
          doi = {10.1051/0004-6361/202451022},
archivePrefix = {arXiv},
       eprint = {2306.06315},
 primaryClass = {astro-ph.CO},
       adsurl = {https://ui.adsabs.harvard.edu/abs/2025A&A...698A.170P}
}

@ARTICLE{Naidoo24,
       author = {{Naidoo}, Krishna and {Lahav}, Ofer},
        title = "{Methods for robustly measuring the minimum spanning tree and other field level statistics from galaxy surveys}",
      journal = {RAS Techniques and Instruments},
         year = 2025,
        month = jan,
       volume = {4},
          eid = {rzaf014},
        pages = {rzaf014},
          doi = {10.1093/rasti/rzaf014},
archivePrefix = {arXiv},
       eprint = {2410.06202},
 primaryClass = {astro-ph.CO},
       adsurl = {https://ui.adsabs.harvard.edu/abs/2025RASTI...4...14N}
}

@ARTICLE{DESI25,
       author = {{Abdul Karim}, M. and {Aguilar}, J. and {Ahlen}, S. and {Alam}, S. and {Allen}, L. and {Prieto}, C. Allende and {Alves}, O. and {Anand}, A. and {Andrade}, U. and {Armengaud}, E. and {Aviles}, A. and {Bailey}, S. and {Baltay}, C. and {Bansal}, P. and {Bault}, A. and {Behera}, J. and {BenZvi}, S. and {Bianchi}, D. and {Blake}, C. and {Brieden}, S. and {Brodzeller}, A. and {Brooks}, D. and {Buckley-Geer}, E. and {Burtin}, E. and {Calderon}, R. and {Canning}, R. and {Rosell}, A. Carnero and {Carrilho}, P. and {Casas}, L. and {Castander}, F.~J. and {Charles}, M. and {Chaussidon}, E. and {Chaves-Montero}, J. and {Chebat}, D. and {Chen}, X. and {Claybaugh}, T. and {Cole}, S. and {Cooper}, A.~P. and {Cuceu}, A. and {Dawson}, K.~S. and {de la Macorra}, A. and {de Mattia}, A. and {Deiosso}, N. and {Della Costa}, J. and {Demina}, R. and {Dey}, A. and {Dey}, B. and {Ding}, Z. and {Doel}, P. and {Edelstein}, J. and {Eisenstein}, D.~J. and {Elbers}, W. and {Fagrelius}, P. and {Fanning}, K. and {Fern{\'a}ndez-Garc{\'\i}a}, E. and {Ferraro}, S. and {Font-Ribera}, A. and {Forero-Romero}, J.~E. and {Frenk}, C.~S. and {Garcia-Quintero}, C. and {Garrison}, L.~H. and {Gazta{\~n}aga}, E. and {Gil-Mar{\'\i}n}, H. and {Gontcho A Gontcho}, S. and {Gonzalez}, D. and {Gonzalez-Morales}, A.~X. and {Gordon}, C. and {Green}, D. and {Gutierrez}, G. and {Guy}, J. and {Hadzhiyska}, B. and {Hahn}, C. and {He}, S. and {Herbold}, M. and {Herrera-Alcantar}, H.~K. and {Ho}, M.-F. and {Honscheid}, K. and {Howlett}, C. and {Huterer}, D. and {Ishak}, M. and {Juneau}, S. and {Kamble}, N.~V. and {Kara{\c{c}}ayl{\i}}, N.~G. and {Kehoe}, R. and {Kent}, S. and {Kim}, A.~G. and {Kirkby}, D. and {Kisner}, T. and {Koposov}, S.~E. and {Kremin}, A. and {Krolewski}, A. and {Lahav}, O. and {Lamman}, C. and {Landriau}, M. and {Lang}, D. and {Lasker}, J. and {Le Goff}, J.~M. and {Le Guillou}, L. and {Leauthaud}, A. and {Levi}, M.~E. and {Li}, Q. and {Li}, T.~S. and {Lodha}, K. and {Lokken}, M. and {Lozano-Rodr{\'\i}guez}, F. and {Magneville}, C. and {Manera}, M. and {Martini}, P. and {Matthewson}, W.~L. and {Meisner}, A. and {Mena-Fern{\'a}ndez}, J. and {Menegas}, A. and {Mergulh{\~a}o}, T. and {Miquel}, R. and {Moustakas}, J. and {Mu{\~n}oz-Guti{\'e}rrez}, A. and {Mu{\~n}oz-Santos}, D. and {Myers}, A.~D. and {Nadathur}, S. and {Naidoo}, K. and {Napolitano}, L. and {Newman}, J.~A. and {Niz}, G. and {Noriega}, H.~E. and {Paillas}, E. and {Palanque-Delabrouille}, N. and {Pan}, J. and {Peacock}, J.~A. and {Pellejero Ibanez}, M. and {Percival}, W.~J. and {P{\'e}rez-Fern{\'a}ndez}, A. and {P{\'e}rez-R{\`a}fols}, I. and {Pieri}, M.~M. and {Poppett}, C. and {Prada}, F. and {Rabinowitz}, D. and {Raichoor}, A. and {Ram{\'\i}rez-P{\'e}rez}, C. and {Rashkovetskyi}, M. and {Ravoux}, C. and {Rich}, J. and {Rocher}, A. and {Rockosi}, C. and {Rohlf}, J. and {Rom{\'a}n-Herrera}, J.~O. and {Ross}, A.~J. and {Rossi}, G. and {Ruggeri}, R. and {Ruhlmann-Kleider}, V. and {Samushia}, L. and {Sanchez}, E. and {Sanders}, N. and {Schlegel}, D. and {Schubnell}, M. and {Seo}, H. and {Shafieloo}, A. and {Sharples}, R. and {Silber}, J. and {Sinigaglia}, F. and {Sprayberry}, D. and {Tan}, T. and {Tarl{\'e}}, G. and {Taylor}, P. and {Turner}, W. and {Ure{\~n}a-L{\'o}pez}, L.~A. and {Vaisakh}, R. and {Valdes}, F. and {Valogiannis}, G. and {Vargas-Maga{\~n}a}, M. and {Verde}, L. and {Walther}, M. and {Weaver}, B.~A. and {Weinberg}, D.~H. and {White}, M. and {Wolfson}, M. and {Y{\`e}che}, C. and {Yu}, J. and {Zaborowski}, E.~A. and {Zarrouk}, P. and {Zhai}, Z. and {Zhang}, H. and {Zhao}, C. and {Zhao}, G.~B. and {Zhou}, R. and {Zou}, H. and {DESI Collaboration}},
        title = "{DESI DR2 results. II. Measurements of baryon acoustic oscillations and cosmological constraints}",
      journal = {\prd},
         year = 2025,
        month = oct,
       volume = {112},
       number = {8},
          eid = {083515},
        pages = {083515},
          doi = {10.1103/tr6y-kpc6},
archivePrefix = {arXiv},
       eprint = {2503.14738},
 primaryClass = {astro-ph.CO},
       adsurl = {https://ui.adsabs.harvard.edu/abs/2025PhRvD.112h3515A}
}

@ARTICLE{FlamingoGeneral23,
       author = {{Schaye}, Joop and {Kugel}, Roi and {Schaller}, Matthieu and {Helly}, John C. and {Braspenning}, Joey and {Elbers}, Willem and {McCarthy}, Ian G. and {van Daalen}, Marcel P. and {Vandenbroucke}, Bert and {Frenk}, Carlos S. and {Kwan}, Juliana and {Salcido}, Jaime and {Bah{\'e}}, Yannick M. and {Borrow}, Josh and {Chaikin}, Evgenii and {Hahn}, Oliver and {Hu{\v{s}}ko}, Filip and {Jenkins}, Adrian and {Lacey}, Cedric G. and {Nobels}, Folkert S.~J.},
        title = "{The FLAMINGO project: cosmological hydrodynamical simulations for large-scale structure and galaxy cluster surveys}",
      journal = {\mnras},
         year = 2023,
        month = dec,
       volume = {526},
       number = {4},
        pages = {4978-5020},
          doi = {10.1093/mnras/stad2419},
archivePrefix = {arXiv},
       eprint = {2306.04024},
 primaryClass = {astro-ph.CO},
       adsurl = {https://ui.adsabs.harvard.edu/abs/2023MNRAS.526.4978S}
}

@ARTICLE{Flamingo_neutrinos25,
       author = {{Elbers}, Willem and {Frenk}, Carlos S. and {Jenkins}, Adrian and {Li}, Baojiu and {Helly}, John C. and {Kugel}, Roi and {Schaller}, Matthieu and {Schaye}, Joop and {Braspenning}, Joey and {Kwan}, Juliana and {McCarthy}, Ian G. and {Salcido}, Jaime and {van Daalen}, Marcel P. and {Vandenbroucke}, Bert and {Pascoli}, Silvia},
        title = "{The FLAMINGO project: the coupling between baryonic feedback and cosmology in light of the S$_{8}$ tension}",
      journal = {\mnras},
         year = 2025,
        month = feb,
       volume = {537},
       number = {2},
        pages = {2160-2178},
          doi = {10.1093/mnras/staf093},
archivePrefix = {arXiv},
       eprint = {2403.12967},
 primaryClass = {astro-ph.CO},
       adsurl = {https://ui.adsabs.harvard.edu/abs/2025MNRAS.537.2160E}
}

@article{HBT_halloes17,
  author = {{Han}, Jiaxin and {Jing}, Y.~P. and {Wang}, Huiyuan and {Wang}, Wenting},
  title = {HBT: Hierarchical Bound-Tracing},
  journal = {\href{https://ui.adsabs.harvard.edu/abs/2017ascl.soft11022H}{Astrophysics Source Code Library}},
  year = 2017,
  month = nov,
  pages = {\href{https://ui.adsabs.harvard.edu/abs/2017ascl.soft11022H}{ascl:1711.022}},
  adsurl = {https://ui.adsabs.harvard.edu/abs/2017ascl.soft11022H}
}

@ARTICLE{PowerofCosmicWeb25Sunseri,
        author = {{Sunseri}, James and {Bayer}, Adrian E. and {Liu}, Jia},
        title = "{Power of the cosmic web}",
      journal = {\prd},
         year = 2025,
        month = sep,
       volume = {112},
       number = {6},
          eid = {063516},
        pages = {063516},
          doi = {10.1103/grx3-hj7w},
archivePrefix = {arXiv},
       eprint = {2503.11778},
 primaryClass = {astro-ph.CO},
       adsurl = {https://ui.adsabs.harvard.edu/abs/2025PhRvD.112f3516S}
}

@ARTICLE{Mummery_2017,
       author = {{Mummery}, Benjamin O. and {McCarthy}, Ian G. and {Bird}, Simeon and {Schaye}, Joop},
        title = "{The separate and combined effects of baryon physics and neutrino free streaming on large-scale structure}",
      journal = {\mnras},
         year = 2017,
        month = oct,
       volume = {471},
       number = {1},
        pages = {227-242},
          doi = {10.1093/mnras/stx1469},
archivePrefix = {arXiv},
       eprint = {1702.02064},
 primaryClass = {astro-ph.CO},
       adsurl = {https://ui.adsabs.harvard.edu/abs/2017MNRAS.471..227M}
}

@ARTICLE{Parroni2021,
       author = {{Parroni}, Carolina and {Tollet}, {\'E}douard and {Cardone}, Vincenzo F. and {Maoli}, Roberto and {Scaramella}, Roberto},
        title = "{Higher-order statistics of shear field via a machine learning approach}",
      journal = {\aap},
         year = 2021,
        month = jan,
       volume = {645},
          eid = {A123},
        pages = {A123},
          doi = {10.1051/0004-6361/202038715},
archivePrefix = {arXiv},
       eprint = {2011.10438},
 primaryClass = {astro-ph.CO},
       adsurl = {https://ui.adsabs.harvard.edu/abs/2021A&A...645A.123P}
}

@ARTICLE{2025EuclidOverview,
       author = {{Euclid Collaboration} and {Mellier}, Y. and {Abdurro'uf} and {Acevedo Barroso}, J.~A. and {Ach{\'u}carro}, A. and {Adamek}, J. and {Adam}, R. and {Addison}, G.~E. and {Aghanim}, N. and {Aguena}, M. and {Ajani}, V. and {Akrami}, Y. and {Al-Bahlawan}, A. and {Alavi}, A. and {Albuquerque}, I.~S. and {Alestas}, G. and {Alguero}, G. and {Allaoui}, A. and {Allen}, S.~W. and {Allevato}, V. and {Alonso-Tetilla}, A.~V. and {Altieri}, B. and {Alvarez-Candal}, A. and {Alvi}, S. and {Amara}, A. and {Amendola}, L. and {Amiaux}, J. and {Andika}, I.~T. and {Andreon}, S. and {Andrews}, A. and {Angora}, G. and {Angulo}, R.~E. and {Annibali}, F. and {Anselmi}, A. and {Anselmi}, S. and {Arcari}, S. and {Archidiacono}, M. and {Aric{\`o}}, G. and {Arnaud}, M. and {Arnouts}, S. and {Asgari}, M. and {Asorey}, J. and {Atayde}, L. and {Atek}, H. and {Atrio-Barandela}, F. and {Aubert}, M. and {Aubourg}, E. and {Auphan}, T. and {Auricchio}, N. and {Aussel}, B. and {Aussel}, H. and {Avelino}, P.~P. and {Avgoustidis}, A. and {Avila}, S. and {Awan}, S. and {Azzollini}, R. and {Baccigalupi}, C. and {Bachelet}, E. and {Bacon}, D. and {Baes}, M. and {Bagley}, M.~B. and {Bahr-Kalus}, B. and {Balaguera-Antolinez}, A. and {Balbinot}, E. and {Balcells}, M. and {Baldi}, M. and {Baldry}, I. and {Balestra}, A. and {Ballardini}, M. and {Ballester}, O. and {Balogh}, M. and {Ba{\~n}ados}, E. and {Barbier}, R. and {Bardelli}, S. and {Baron}, M. and {Barreiro}, T. and {Barrena}, R. and {Barriere}, J. -C. and {Barros}, B.~J. and {Barthelemy}, A. and {Bartolo}, N. and {Basset}, A. and {Battaglia}, P. and {Battisti}, A.~J. and {Baugh}, C.~M. and {Baumont}, L. and {Bazzanini}, L. and {Beaulieu}, J. -P. and {Beckmann}, V. and {Belikov}, A.~N. and {Bel}, J. and {Bellagamba}, F. and {Bella}, M. and {Bellini}, E. and {Benabed}, K. and {Bender}, R. and {Benevento}, G. and {Bennett}, C.~L. and {Benson}, K. and {Bergamini}, P. and {Bermejo-Climent}, J.~R. and {Bernardeau}, F. and {Bertacca}, D. and {Berthe}, M. and {Berthier}, J. and {Bethermin}, M. and {Beutler}, F. and {Bevillon}, C. and {Bhargava}, S. and {Bhatawdekar}, R. and {Bianchi}, D. and {Bisigello}, L. and {Biviano}, A. and {Blake}, R.~P. and {Blanchard}, A. and {Blazek}, J. and {Blot}, L. and {Bosco}, A. and {Bodendorf}, C. and {Boenke}, T. and {B{\"o}hringer}, H. and {Boldrini}, P. and {Bolzonella}, M. and {Bonchi}, A. and {Bonici}, M. and {Bonino}, D. and {Bonino}, L. and {Bonvin}, C. and {Bon}, W. and {Booth}, J.~T. and {Borgani}, S. and {Borlaff}, A.~S. and {Borsato}, E. and {Bose}, B. and {Botticella}, M.~T. and {Boucaud}, A. and {Bouche}, F. and {Boucher}, J.~S. and {Boutigny}, D. and {Bouvard}, T. and {Bouwens}, R. and {Bouy}, H. and {Bowler}, R.~A.~A. and {Bozza}, V. and {Bozzo}, E. and {Branchini}, E. and {Brando}, G. and {Brau-Nogue}, S. and {Brekke}, P. and {Bremer}, M.~N. and {Brescia}, M. and {Breton}, M. -A. and {Brinchmann}, J. and {Brinckmann}, T. and {Brockley-Blatt}, C. and {Brodwin}, M. and {Brouard}, L. and {Brown}, M.~L. and {Bruton}, S. and {Bucko}, J. and {Buddelmeijer}, H. and {Buenadicha}, G. and {Buitrago}, F. and {Burger}, P. and {Burigana}, C. and {Busillo}, V. and {Busonero}, D. and {Cabanac}, R. and {Cabayol-Garcia}, L. and {Cagliari}, M.~S. and {Caillat}, A. and {Caillat}, L. and {Calabrese}, M. and {Calabro}, A. and {Calderone}, G. and {Calura}, F. and {Camacho Quevedo}, B. and {Camera}, S. and {Campos}, L. and {Ca{\~n}as-Herrera}, G. and {Candini}, G.~P. and {Cantiello}, M. and {Capobianco}, V. and {Cappellaro}, E. and {Cappelluti}, N. and {Cappi}, A. and {Caputi}, K.~I. and {Cara}, C. and {Carbone}, C. and {Cardone}, V.~F. and {Carella}, E. and {Carlberg}, R.~G. and {Carle}, M. and {Carminati}, L. and {Caro}, F. and {Carrasco}, J.~M. and {Carretero}, J. and {Carrilho}, P. and {Carron Duque}, J. and {Carry}, B.},
        title = "{Euclid: I. Overview of the Euclid mission}",
      journal = {\aap},
         year = 2025,
        month = may,
       volume = {697},
          eid = {A1},
        pages = {A1},
          doi = {10.1051/0004-6361/202450810},
archivePrefix = {arXiv},
       eprint = {2405.13491},
 primaryClass = {astro-ph.CO},
       adsurl = {https://ui.adsabs.harvard.edu/abs/2025A&A...697A...1E}
}

@ARTICLE{2022DESI_overview,
       author = {{DESI Collaboration} and {Abareshi}, B. and {Aguilar}, J. and {Ahlen}, S. and {Alam}, Shadab and {Alexander}, David M. and {Alfarsy}, R. and {Allen}, L. and {Allende Prieto}, C. and {Alves}, O. and {Ameel}, J. and {Armengaud}, E. and {Asorey}, J. and {Aviles}, Alejandro and {Bailey}, S. and {Balaguera-Antol{\'\i}nez}, A. and {Ballester}, O. and {Baltay}, C. and {Bault}, A. and {Beltran}, S.~F. and {Benavides}, B. and {BenZvi}, S. and {Berti}, A. and {Besuner}, R. and {Beutler}, Florian and {Bianchi}, D. and {Blake}, C. and {Blanc}, P. and {Blum}, R. and {Bolton}, A. and {Bose}, S. and {Bramall}, D. and {Brieden}, S. and {Brodzeller}, A. and {Brooks}, D. and {Brownewell}, C. and {Buckley-Geer}, E. and {Cahn}, R.~N. and {Cai}, Z. and {Canning}, R. and {Capasso}, R. and {Carnero Rosell}, A. and {Carton}, P. and {Casas}, R. and {Castander}, F.~J. and {Cervantes-Cota}, J.~L. and {Chabanier}, S. and {Chaussidon}, E. and {Chuang}, C. and {Circosta}, C. and {Cole}, S. and {Cooper}, A.~P. and {da Costa}, L. and {Cousinou}, M. -C. and {Cuceu}, A. and {Davis}, T.~M. and {Dawson}, K. and {de la Cruz-Noriega}, R. and {de la Macorra}, A. and {de Mattia}, A. and {Della Costa}, J. and {Demmer}, P. and {Derwent}, M. and {Dey}, A. and {Dey}, B. and {Dhungana}, G. and {Ding}, Z. and {Dobson}, C. and {Doel}, P. and {Donald-McCann}, J. and {Donaldson}, J. and {Douglass}, K. and {Duan}, Y. and {Dunlop}, P. and {Edelstein}, J. and {Eftekharzadeh}, S. and {Eisenstein}, D.~J. and {Enriquez-Vargas}, M. and {Escoffier}, S. and {Evatt}, M. and {Fagrelius}, P. and {Fan}, X. and {Fanning}, K. and {Fawcett}, V.~A. and {Ferraro}, S. and {Ereza}, J. and {Flaugher}, B. and {Font-Ribera}, A. and {Forero-Romero}, J.~E. and {Frenk}, C.~S. and {Fromenteau}, S. and {G{\"a}nsicke}, B.~T. and {Garcia-Quintero}, C. and {Garrison}, L. and {Gazta{\~n}aga}, E. and {Gerardi}, F. and {Gil-Mar{\'\i}n}, H. and {Gontcho A Gontcho}, S. and {Gonzalez-Morales}, Alma X. and {Gonzalez-de-Rivera}, G. and {Gonzalez-Perez}, V. and {Gordon}, C. and {Graur}, O. and {Green}, D. and {Grove}, C. and {Gruen}, D. and {Gutierrez}, G. and {Guy}, J. and {Hahn}, C. and {Harris}, S. and {Herrera}, D. and {Herrera-Alcantar}, Hiram K. and {Honscheid}, K. and {Howlett}, C. and {Huterer}, D. and {Ir{\v{s}}i{\v{c}}}, V. and {Ishak}, M. and {Jelinsky}, P. and {Jiang}, L. and {Jimenez}, J. and {Jing}, Y.~P. and {Joyce}, R. and {Jullo}, E. and {Juneau}, S. and {Kara{\c{c}}ayl{\i}}, N.~G. and {Karamanis}, M. and {Karcher}, A. and {Karim}, T. and {Kehoe}, R. and {Kent}, S. and {Kirkby}, D. and {Kisner}, T. and {Kitaura}, F. and {Koposov}, S.~E. and {Kov{\'a}cs}, A. and {Kremin}, A. and {Krolewski}, Alex and {L'Huillier}, B. and {Lahav}, O. and {Lambert}, A. and {Lamman}, C. and {Lan}, Ting-Wen and {Landriau}, M. and {Lane}, S. and {Lang}, D. and {Lange}, J.~U. and {Lasker}, J. and {Le Guillou}, L. and {Leauthaud}, A. and {Le Van Suu}, A. and {Levi}, Michael E. and {Li}, T.~S. and {Magneville}, C. and {Manera}, M. and {Manser}, Christopher J. and {Marshall}, B. and {Martini}, Paul and {McCollam}, W. and {McDonald}, P. and {Meisner}, Aaron M. and {Mena-Fern{\'a}ndez}, J. and {Meneses-Rizo}, J. and {Mezcua}, M. and {Miller}, T. and {Miquel}, R. and {Montero-Camacho}, P. and {Moon}, J. and {Moustakas}, J. and {Mueller}, E. and {Mu{\~n}oz-Guti{\'e}rrez}, Andrea and {Myers}, Adam D. and {Nadathur}, S. and {Najita}, J. and {Napolitano}, L. and {Neilsen}, E. and {Newman}, Jeffrey A. and {Nie}, J.~D. and {Ning}, Y. and {Niz}, G. and {Norberg}, P. and {Noriega}, Hern{\'a}n E. and {O'Brien}, T. and {Obuljen}, A. and {Palanque-Delabrouille}, N. and {Palmese}, A. and {Zhiwei}, P. and {Pappalardo}, D. and {PENG}, X. and {Percival}, W.~J. and {Perruchot}, S. and {Pogge}, R. and {Poppett}, C. and {Porredon}, A. and {Prada}, F. and {Prochaska}, J. and {Pucha}, R. and {P{\'e}rez-Fern{\'a}ndez}, A. and {P{\'e}rez-R{\`a}fols}, I. and {Rabinowitz}, D. and {Raichoor}, A.},
        title = "{Overview of the Instrumentation for the Dark Energy Spectroscopic Instrument}",
      journal = {\aj},
         year = 2022,
        month = nov,
       volume = {164},
       number = {5},
          eid = {207},
        pages = {207},
          doi = {10.3847/1538-3881/ac882b},
archivePrefix = {arXiv},
       eprint = {2205.10939},
 primaryClass = {astro-ph.IM},
       adsurl = {https://ui.adsabs.harvard.edu/abs/2022AJ....164..207D}
}

@ARTICLE{2019LSST_Overview,
       author = {{Ivezi{\'c}}, {\v{Z}}eljko and {Kahn}, Steven M. and {Tyson}, J. Anthony and {Abel}, Bob and {Acosta}, Emily and {Allsman}, Robyn and {Alonso}, David and {AlSayyad}, Yusra and {Anderson}, Scott F. and {Andrew}, John and {Angel}, James Roger P. and {Angeli}, George Z. and {Ansari}, Reza and {Antilogus}, Pierre and {Araujo}, Constanza and {Armstrong}, Robert and {Arndt}, Kirk T. and {Astier}, Pierre and {Aubourg}, {\'E}ric and {Auza}, Nicole and {Axelrod}, Tim S. and {Bard}, Deborah J. and {Barr}, Jeff D. and {Barrau}, Aurelian and {Bartlett}, James G. and {Bauer}, Amanda E. and {Bauman}, Brian J. and {Baumont}, Sylvain and {Bechtol}, Ellen and {Bechtol}, Keith and {Becker}, Andrew C. and {Becla}, Jacek and {Beldica}, Cristina and {Bellavia}, Steve and {Bianco}, Federica B. and {Biswas}, Rahul and {Blanc}, Guillaume and {Blazek}, Jonathan and {Blandford}, Roger D. and {Bloom}, Josh S. and {Bogart}, Joanne and {Bond}, Tim W. and {Booth}, Michael T. and {Borgland}, Anders W. and {Borne}, Kirk and {Bosch}, James F. and {Boutigny}, Dominique and {Brackett}, Craig A. and {Bradshaw}, Andrew and {Brandt}, William Nielsen and {Brown}, Michael E. and {Bullock}, James S. and {Burchat}, Patricia and {Burke}, David L. and {Cagnoli}, Gianpietro and {Calabrese}, Daniel and {Callahan}, Shawn and {Callen}, Alice L. and {Carlin}, Jeffrey L. and {Carlson}, Erin L. and {Chandrasekharan}, Srinivasan and {Charles-Emerson}, Glenaver and {Chesley}, Steve and {Cheu}, Elliott C. and {Chiang}, Hsin-Fang and {Chiang}, James and {Chirino}, Carol and {Chow}, Derek and {Ciardi}, David R. and {Claver}, Charles F. and {Cohen-Tanugi}, Johann and {Cockrum}, Joseph J. and {Coles}, Rebecca and {Connolly}, Andrew J. and {Cook}, Kem H. and {Cooray}, Asantha and {Covey}, Kevin R. and {Cribbs}, Chris and {Cui}, Wei and {Cutri}, Roc and {Daly}, Philip N. and {Daniel}, Scott F. and {Daruich}, Felipe and {Daubard}, Guillaume and {Daues}, Greg and {Dawson}, William and {Delgado}, Francisco and {Dellapenna}, Alfred and {de Peyster}, Robert and {de Val-Borro}, Miguel and {Digel}, Seth W. and {Doherty}, Peter and {Dubois}, Richard and {Dubois-Felsmann}, Gregory P. and {Durech}, Josef and {Economou}, Frossie and {Eifler}, Tim and {Eracleous}, Michael and {Emmons}, Benjamin L. and {Fausti Neto}, Angelo and {Ferguson}, Henry and {Figueroa}, Enrique and {Fisher-Levine}, Merlin and {Focke}, Warren and {Foss}, Michael D. and {Frank}, James and {Freemon}, Michael D. and {Gangler}, Emmanuel and {Gawiser}, Eric and {Geary}, John C. and {Gee}, Perry and {Geha}, Marla and {Gessner}, Charles J.~B. and {Gibson}, Robert R. and {Gilmore}, D. Kirk and {Glanzman}, Thomas and {Glick}, William and {Goldina}, Tatiana and {Goldstein}, Daniel A. and {Goodenow}, Iain and {Graham}, Melissa L. and {Gressler}, William J. and {Gris}, Philippe and {Guy}, Leanne P. and {Guyonnet}, Augustin and {Haller}, Gunther and {Harris}, Ron and {Hascall}, Patrick A. and {Haupt}, Justine and {Hernandez}, Fabio and {Herrmann}, Sven and {Hileman}, Edward and {Hoblitt}, Joshua and {Hodgson}, John A. and {Hogan}, Craig and {Howard}, James D. and {Huang}, Dajun and {Huffer}, Michael E. and {Ingraham}, Patrick and {Innes}, Walter R. and {Jacoby}, Suzanne H. and {Jain}, Bhuvnesh and {Jammes}, Fabrice and {Jee}, M. James and {Jenness}, Tim and {Jernigan}, Garrett and {Jevremovi{\'c}}, Darko and {Johns}, Kenneth and {Johnson}, Anthony S. and {Johnson}, Margaret W.~G. and {Jones}, R. Lynne and {Juramy-Gilles}, Claire and {Juri{\'c}}, Mario and {Kalirai}, Jason S. and {Kallivayalil}, Nitya J. and {Kalmbach}, Bryce and {Kantor}, Jeffrey P. and {Karst}, Pierre and {Kasliwal}, Mansi M. and {Kelly}, Heather and {Kessler}, Richard and {Kinnison}, Veronica and {Kirkby}, David and {Knox}, Lloyd and {Kotov}, Ivan V. and {Krabbendam}, Victor L. and {Krughoff}, K. Simon and {Kub{\'a}nek}, Petr and {Kuczewski}, John and {Kulkarni}, Shri and {Ku}, John and {Kurita}, Nadine R. and {Lage}, Craig S. and {Lambert}, Ron and {Lange}, Travis and {Langton}, J. Brian and {Le Guillou}, Laurent and {Levine}, Deborah and {Liang}, Ming and {Lim}, Kian-Tat and {Lintott}, Chris J. and {Long}, Kevin E. and {Lopez}, Margaux and {Lotz}, Paul J. and {Lupton}, Robert H. and {Lust}, Nate B. and {MacArthur}, Lauren A. and {Mahabal}, Ashish and {Mandelbaum}, Rachel and {Markiewicz}, Thomas W. and {Marsh}, Darren S. and {Marshall}, Philip J. and {Marshall}, Stuart and {May}, Morgan and {McKercher}, Robert and {McQueen}, Michelle and {Meyers}, Joshua and {Migliore}, Myriam and {Miller}, Michelle and {Mills}, David J.},
        title = "{LSST: From Science Drivers to Reference Design and Anticipated Data Products}",
      journal = {\apj},
         year = 2019,
        month = mar,
       volume = {873},
       number = {2},
          eid = {111},
        pages = {111},
          doi = {10.3847/1538-4357/ab042c},
archivePrefix = {arXiv},
       eprint = {0805.2366},
 primaryClass = {astro-ph},
       adsurl = {https://ui.adsabs.harvard.edu/abs/2019ApJ...873..111I}
}

@ARTICLE{2025FlamingoBaryonFeedback,
       author = {{Schaller}, Matthieu and {Schaye}, Joop and {Kugel}, Roi and {Broxterman}, Jeger C. and {van Daalen}, Marcel P.},
        title = "{The FLAMINGO project: baryon effects on the matter power spectrum}",
      journal = {\mnras},
         year = 2025,
        month = may,
       volume = {539},
       number = {2},
        pages = {1337-1351},
          doi = {10.1093/mnras/staf569},
archivePrefix = {arXiv},
       eprint = {2410.17109},
 primaryClass = {astro-ph.CO},
       adsurl = {https://ui.adsabs.harvard.edu/abs/2025MNRAS.539.1337S}
}

@ARTICLE{1998WeigheingNeutGalSur,
       author = {{Hu}, Wayne and {Eisenstein}, Daniel J. and {Tegmark}, Max},
        title = "{Weighing Neutrinos with Galaxy Surveys}",
      journal = {\prl},
         year = 1998,
        month = jun,
       volume = {80},
       number = {24},
        pages = {5255-5258},
          doi = {10.1103/PhysRevLett.80.5255},
archivePrefix = {arXiv},
       eprint = {astro-ph/9712057},
 primaryClass = {astro-ph},
       adsurl = {https://ui.adsabs.harvard.edu/abs/1998PhRvL..80.5255H}
}

@ARTICLE{2022Wavelet,
       author = {{Valogiannis}, Georgios and {Dvorkin}, Cora},
        title = "{Going beyond the galaxy power spectrum: An analysis of BOSS data with wavelet scattering transforms}",
      journal = {\prd},
         year = 2022,
        month = nov,
       volume = {106},
       number = {10},
          eid = {103509},
        pages = {103509},
          doi = {10.1103/PhysRevD.106.103509},
archivePrefix = {arXiv},
       eprint = {2204.13717},
 primaryClass = {astro-ph.CO},
       adsurl = {https://ui.adsabs.harvard.edu/abs/2022PhRvD.106j3509V}
}

@ARTICLE{2025Minkowski,
       author = {{Liu}, Wei and {Paillas}, Enrique and {Cuesta-Lazaro}, Carolina and {Valogiannis}, Georgios and {Fang}, Wenjuan},
        title = "{Cosmological constraints from the Minkowski functionals of the BOSS CMASS galaxy sample}",
      journal = {\jcap},
         year = 2025,
        month = may,
       volume = {2025},
       number = {5},
          eid = {064},
        pages = {064},
          doi = {10.1088/1475-7516/2025/05/064},
archivePrefix = {arXiv},
       eprint = {2501.01698},
 primaryClass = {astro-ph.CO},
       adsurl = {https://ui.adsabs.harvard.edu/abs/2025JCAP...05..064L}
}

@ARTICLE{2023densitysplit1,
       author = {{Paillas}, Enrique and {Cuesta-Lazaro}, Carolina and {Zarrouk}, Pauline and {Cai}, Yan-Chuan and {Percival}, Will J. and {Nadathur}, Seshadri and {Pinon}, Mathilde and {de Mattia}, Arnaud and {Beutler}, Florian},
        title = "{Constraining {\ensuremath{\nu}}{\ensuremath{\Lambda}}CDM with density-split clustering}",
      journal = {\mnras},
         year = 2023,
        month = jun,
       volume = {522},
       number = {1},
        pages = {606-625},
          doi = {10.1093/mnras/stad1017},
archivePrefix = {arXiv},
       eprint = {2209.04310},
 primaryClass = {astro-ph.CO},
       adsurl = {https://ui.adsabs.harvard.edu/abs/2023MNRAS.522..606P}
}

@ARTICLE{2024densitysplit2,
       author = {{Paillas}, Enrique and {Cuesta-Lazaro}, Carolina and {Percival}, Will J. and {Nadathur}, Seshadri and {Cai}, Yan-Chuan and {Yuan}, Sihan and {Beutler}, Florian and {de Mattia}, Arnaud and {Eisenstein}, Daniel J. and {Forero-Sanchez}, Daniel and {Padilla}, Nelson and {Pinon}, Mathilde and {Ruhlmann-Kleider}, Vanina and {S{\'a}nchez}, Ariel G. and {Valogiannis}, Georgios and {Zarrouk}, Pauline},
        title = "{Cosmological constraints from density-split clustering in the BOSS CMASS galaxy sample}",
      journal = {\mnras},
         year = 2024,
        month = jun,
       volume = {531},
       number = {1},
        pages = {898-918},
          doi = {10.1093/mnras/stae1118},
archivePrefix = {arXiv},
       eprint = {2309.16541},
 primaryClass = {astro-ph.CO},
       adsurl = {https://ui.adsabs.harvard.edu/abs/2024MNRAS.531..898P}
}

@ARTICLE{2022NpointStats,
       author = {{Philcox}, Oliver H.~E. and {Slepian}, Zachary},
        title = "{Efficient computation of N-point correlation functions in D dimensions}",
      journal = {Proceedings of the National Academy of Science},
         year = 2022,
        month = aug,
       volume = {119},
       number = {33},
          eid = {e2111366119},
        pages = {e2111366119},
          doi = {10.1073/pnas.2111366119},
archivePrefix = {arXiv},
       eprint = {2106.10278},
 primaryClass = {astro-ph.IM},
       adsurl = {https://ui.adsabs.harvard.edu/abs/2022PNAS..11911366P}
}

@ARTICLE{Brandbyge10,
       author = {{Brandbyge}, Jacob and {Hannestad}, Steen and {Haugb{\o}lle}, Troels and {Wong}, Yvonne Y.~Y.},
        title = "{Neutrinos in non-linear structure formation {\textemdash} the effect on halo properties}",
      journal = {\jcap},
         year = 2010,
        month = sep,
       volume = {2010},
       number = {9},
          eid = {014},
        pages = {014},
          doi = {10.1088/1475-7516/2010/09/014},
archivePrefix = {arXiv},
       eprint = {1004.4105},
 primaryClass = {astro-ph.CO},
       adsurl = {https://ui.adsabs.harvard.edu/abs/2010JCAP...09..014B}
}

\appendix

\section{Exploring varied cosmology}

\begin{table*}
    \centering
    \caption{Cosmological parameters of the \textit{Fix} and \textit{Var} FLAMINGO 
    simulations at $M_\nu = 0.24\,\text{eV}$ used in this appendix 
    \citep{FlamingoGeneral23, Flamingo_neutrinos25}. Columns are as 
    described in Table~\ref{tab:flamingo}.}
    \label{tab:flamingo_var}
    \begin{tabular}{lcccccccccc}
        \hline
        Simulation Cosmology & $h$ & $\Omega_\text{m}$ & $\Omega_\text{c}$ & $\Omega_\text{b}$ & $M_\nu$ & $\sigma_8$ & $10^9 A_\text{s}$ & $n_\text{s}$ \\
        \hline
        {\footnotesize Planck$\nu$0.24Fix} & $0.673$ & $0.316$ & $0.261$ & $0.0494$ & $\SI{0.24}{\eV}$ & $0.769$ & $2.101$ & $0.966$ \\
        {\footnotesize Planck$\nu$0.24Var} & $0.662$ & $0.328$ & $0.263$ & $0.0510$ & $\SI{0.24}{\eV}$ & $0.772$ & $2.109$ & $0.968$ \\
        \hline
    \end{tabular}
\end{table*}

\begin{figure*}
    \centering
    \includegraphics[width=\textwidth]{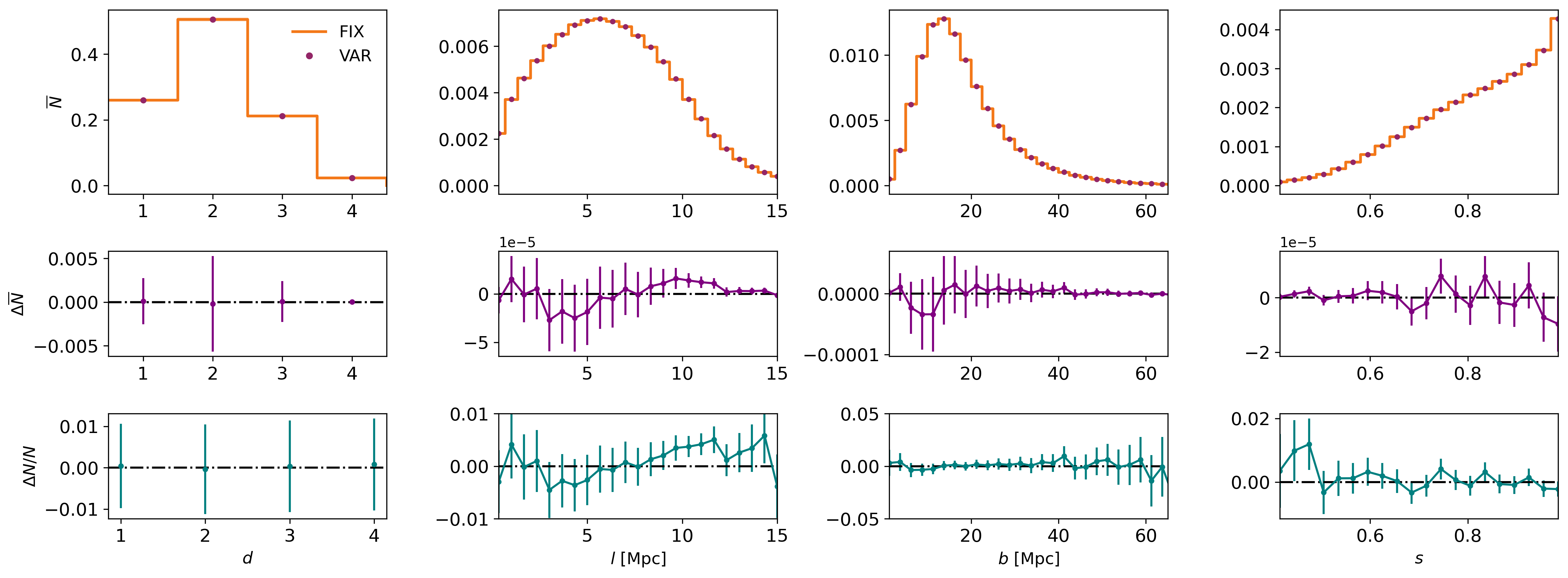}
    \caption{A comparison of the MST statistics between the \textit{Fix} and \textit{Var} 
    cosmologies at $M_\nu = 0.24\,\text{eV}$, analogous to Fig.~\ref{fig:mst_comp}. 
    The top panels show the distributions of degrees ($d$), edge lengths ($l$), branch lengths ($b$), and branch shapes ($s$). The middle 
    panels show the normalised differences between the two ($\Delta\bar{N} = \bar{N}_{VAR} - \bar{N}_{FIX}$), 
    and the bottom panels show the fractional differences 
    ($\Delta N/N$). The differences are at the sub-percent level across all statistics, within the jackknife uncertainties, confirming that the MST signal is robust to the choice of \textit{Fix} 
    versus \textit{Var} cosmology.}
    \label{fig:mst_appendix_varied}
\end{figure*}

The simulations used in this paper adopt fixed cosmological parameters while varying the neutrino mass (\textit{Fix} cosmology), meaning parameters such as $\sigma_8$ and $\Omega_\text{c}$ are not adjusted to remain consistent with CMB observations as $M_\nu$ increases (see Table \ref{tab:flamingo}). In reality, increasing the neutrino mass while holding $A_s$ fixed leads to a lower $\sigma_8$, and a truly CMB-consistent cosmology would compensate by adjusting other parameters accordingly. The FLAMINGO suite includes simulations with a varying cosmology (\textit{Var}) for $M_\nu = 0.24\,\text{eV}$, in which the other cosmological parameters are simultaneously varied to retain a good fit to CMB and other observational constraints (see Table \ref{tab:flamingo_var} for cosmologies). We use this simulation to test to what extent our results are affected by this choice.

In Fig.~\ref{fig:mst_appendix_varied} we show a comparison of the MST statistics between the \textit{Fix} and \textit{Var} cosmologies, both DMO simulations and with a neutrino mass at $M_\nu = 0.24\,\text{eV}$. The top two panels are analogous to Fig.~\ref{fig:mst_comp}, and we add a third panel that shows the fractional difference in the statistics. The differences between the two are at the sub-percent level across all four MST statistics, within the jackknife uncertainties. This suggests that the MST signal is driven primarily by the neutrino mass itself, rather than by the accompanying shifts in other cosmological parameters.

The main reason we adopt the \textit{Fix} cosmology throughout this paper is that it is the only variant available at $M_\nu = 0.48\,\text{eV}$, which we need to study the neutrino mass dependence across a broader range. Given the negligible difference seen here between \textit{Fix} and \textit{Var} at $M_\nu = 0.24\,\text{eV}$, we are confident that the trends and conclusions presented in the main body of this paper are robust to this choice.

%%%%%%%%%%%%%%%%%%%%%%%%%%%%%%%%%%%%%%%%%%%%%%%%%%

%%%%%%%%%%%%%%%%% APPENDICES %%%%%%%%%%%%%%%%%%%%%

%\appendix


%%%%%%%%%%%%%%%%%%%%%%%%%%%%%%%%%%%%%%%%%%%%%%%%%%

% Don't change these lines
\bsp	% typesetting comment
\label{lastpage}
\end{document}